\documentclass[preprint,12pt,3p]{elsarticle}

 \usepackage{graphicx}

\usepackage{amssymb}
\usepackage{amsmath}
\usepackage{xcolor}
\usepackage{lineno}
\usepackage{float}
\usepackage{subcaption}
\usepackage{subfloat}
\floatstyle{plaintop}
\restylefloat{table}

\journal{Applied Ocean Research}

\begin{document}

\begin{frontmatter}

\title{Down-scale marine hydrodynamic analysis at the Norwegian coast - the NORA-SARAH open framework}

\author[label1]{Widar Weizhi Wang\corref{cor1}}

\address[label1]{Norwegian University of Science and Technology \\ H{\o}gskoleringen 7A, 7491 Trondheim, Norway}
\address[label2]{Norwegian Meteorological Institute, All\'{e}gaten 70, 5007 Bergen, Norway}
\address[label3]{SINTEF Ocean, Jonsvannsveien 82, 7050 Trondheim, Norway}

\cortext[cor1]{I am corresponding author}
\ead{widar.w.wang@ntnu.no}

\author[label2]{Konstantinos Christakos}
\ead{konstantinosc@met.no}

\author[label3]{Csaba P\'{a}kozdi}
\ead{csaba.pakozdi@sintef.no}

\author[label1]{Hans Bihs}
\ead{hans.bihs@ntnu.no}


\begin{abstract}

Offshore wave studies often assume Gaussian processes and homogeneous wave fields. However, as waves approach the shoreline, complex coastal topo-bathymetry induces transformations such as shoaling, refraction, diffraction, reflection, and breaking, leading to increased nonlinearity and site-specific wave characteristics. This complexity necessitates detailed site-specific studies for coastal infrastructure design and blue economy planning.

This work presents a downscaling procedure for analyzing wave-structure interactions from offshore metocean conditions. The open-access NORA3 and NORA10EI hindcast databases define offshore seastates, which are propagated to nearshore regions using the phase-averaged wave model SWAN. The outputs inform phase-resolving simulations with the fully nonlinear potential flow solver REEF3D::FNPF, incorporating an Arbitrary Eulerian-Lagrangian (ALE) method to compute wave forces via Morison’s formulation and to screen for extreme events. Extreme wave loads are further examined using the fully viscous Navier-Stokes solver REEF3D::CFD. A one-way hydrodynamic coupling (HDC) between the potential flow and viscous solvers ensures accurate information transfer.

The proposed NORA-SARAH framework, integrating NORA databases with SWAN-REEF3D-ALE-HDC, offers a robust approach for complex coastal environments. A case study in Southern Norway demonstrates its advantages over traditional significant wave height ($H_s$)-based or phase-averaged modeling-based practices, highlighting the necessity of this downscaling method.
\end{abstract}


\end{frontmatter}

\section{Introduction}
\label{sec1:introduction}




The study of offshore wave dynamics has traditionally relied on the assumption of Gaussian processes, treating wave fields as both homogeneous and ergodic. For offshore engineering applications, the characterization of sea states is often conducted using well-established spectral formulations, such as the Pierson-Moskowitz spectrum \cite{Pierson1964} for fully developed seas and the JONSWAP spectrum \cite{Hasselmann_JONSWAP_1973} for fetch-limited, developing wave conditions. These spectral models effectively describe deep-water waves; however, as waves propagate towards the coast and interact with complex coastal topo-bathymetry, a series of inter-correlated wave transformations occur, such as shoaling, refraction, diffraction, reflection and breaking. These wave transformations introduce additional nonlinearity and render the wave fields inhomogeneous. Observations from Norwegian coastal waters reveal pronounced inhomogeneities in the wave field, as evidenced by field measurements \cite{CHENG_2021_fjord_inhomogeneity, CHRISTAKOS_2022_fjord_inhomogeneity} and laboratory experiments \cite{LAFLECHE_2024_exp_inhomogeneity}.  Though corrections can be made on the established spectra to account for the shallow water effects such as the TMA spectrum \cite{Hughes_TMA}, it is challenging to formulate a coastal wave spectrum considering the inhomogeneous and nonlinear complex wave fields nearshore. Consequently, coastal wave conditions tend to be highly site-specific, lacking a universally applicable formulation. This limitation poses significant challenges for marine structural design and the planning of blue economy activities, necessitating detailed site-specific investigations to ensure reliable engineering assessments.

One of the first fundamental steps of the site-specific studies is to analyze the meteorology and oceanography (metocean) data at the site of interest. It has been challenging to obtain such data in the past, but increasing efforts at both national and international levels have significantly improved data availability through open-access initiatives. In Norway, the Norwegian Meteorological Institute (MET) has played a pivotal role in producing comprehensive metocean datasets that are freely accessible to both industry and the public. Among these datasets, the NORA10 hindcast database provides long-term wind and wave hindcasts for the Norwegian Sea, the North Sea, and the Barents Sea, covering the period from 1957 to 2022 with a spatial resolution of approximately 10 km. This dataset is based on atmospheric downscaling of the ERA-40 reanalysis \citep{Reistad_2011, Aarnes_2012, Furevik_2012}, and can be accessed through direct requests or via the revised open-access NORA10EI database \citep{Haakenstad_2020_NORA10EI}. More recently, the NORA3 dataset \citep{Haakenstad_2021_NORA3_Wind, Breivik_2022_NORA3_Wave} has been introduced, offering improved resolution (3 km horizontal spacing), an expanded spatial domain encompassing the Nordic Seas and the Arctic Ocean, and up-to-date wave and wind information. The accessibility of these databases has been further enhanced by the DNORA API \citep{Christakos_2023_DNORA}, enabling seamless integration with open-source wave models such as SWAN \citep{Booij_1999} and REEF3D \cite{bihs2016cf, Wang_2020_compariosn}. Despite these advancements, a significant scale gap persists between metocean databases and the resolution requirements of coastal engineering applications. While metocean datasets provide regional wave characteristics at resolutions between 3 and 10 km, coastal engineering sites themselves can be smaller than 1 ${km}^2$. Consequently, crucial hydrodynamic processes that influence wave-structure interactions remain unresolved, as they are typically confined within a single grid cell or data point of the metocean dataset.

To address this scale gap, spectral wave models such as SWAN \citep{Booij_1999} serve as an intermediate tool, translating large-scale metocean conditions into coastal wave characteristics. SWAN's computational efficiency and capability to simulate wind-wave generation, energy dissipation, shoaling, and nonlinear wave-wave interactions make it an essential component of coastal wave modeling. However, despite its strengths, the underlying assumptions in spectral models limit their ability to capture strongly nonlinear wave transformations and pronounced diffraction effects \citep{Holthuijsen_2003}. As a result, phase-resolving models become necessary for accurately simulating complex coastal wave fields.

A variety of phase-resolving models have been developed, each with distinct mathematical formulations and assumptions. Many models rely on depth-averaged approaches based on shallow-water equations (SWE), including Boussinesq-type models \cite{madsen1991, MADSEN1992183, nwogu1993, Madsen1998}, non-hydrostatic multi-layer models \cite{Lynett_2004, Stelling_2003, zijlema2005, zijlema2008, zijlema_2011}, and quadratic non-hydrostatic pressure models \cite{Jeschke2017, Wang_2020_IJNMF}. However, the shallow-water assumption is not universally applicable, particularly in coastal regions with abrupt bathymetric variations, such as the Norwegian fjords \citep{Wang_2020_PhD}. The inherent limitation of SWE-based models is the ability to represent the deep water dispersion relations. In order to achieve better dispersion, Boussinesq models employ higher-order explicit non-hydrostatic pressure terms \cite{gobbi_kirby_wei_2000, wei_1995}. Despite the achievement of modeling waves accurately up to the non-dimensional water depth to wavelength $kd=40$ presented by \cite{madsen2002}, the multiple expansions lead to a large set of equations and high-order derivatives which limit the numerical stability of the models. The multi-layer models can also achieve much-improved dispersion relation with an increasing number of vertical layers, but often at the cost of significantly increased computational time \cite{Monteban2016}. 

To overcome these limitations, potential-flow-based models, including boundary element methods (BEM) \cite{Grilli_1994, Grilli_2001}, high-order spectral (HOS) methods \cite{Ducrozet_2012, BONNEFOY200633, BONNEFOY2006121, Ducrozet_2016, Raoult_2016_194, Yates_2015}, and fully nonlinear potential flow (FNPF) models \cite{Li1997235, Bingham2007, EngsigKarup_2009, Bihs_2020_FNPF}, provide more accurate descriptions of wave dispersion. For large-scale coastal areas with irregular boundaries, the fully populated unsymmetrical matrix in a BEM model is computationally demanding and makes it challenging to implement parallel computation techniques suitable for large domains. In recent developments, the HOS model has improved its ability to represent wave propagation over varying bathymetry and breaking waves \cite{ZhangBenoit2019, Simon2019} though practical engineering applications are yet to be shown. The finite difference method (FDM)-based FNPF model OceanWaves3D \cite{EngsigKarup_2009} uses a surface-and-bottom-following $\sigma$-grid to represent the varying bathymetry and an adaptive curvilinear grid to represent the irregular coastlines. The model is thus very versatile for coastal applications, especially given the recent GPU acceleration \cite{EngsigKarup_2012}. However, the generation of a curvilinear grid is case-dependent and can be time-consuming. To further the versatility, the REEF3D::FNPF model \cite{Wang_2022_AOR_FNPF} introduces an innovative level-set-based coastline algorithm that ensures a flexible and efficient representation of coastal boundaries while maintaining a straightforward structured rectilinear horizontal grid. REEF3D::FNPF has also included the steepness-induced and depth-induced breaking wave algorithms to identify breaking waves and represent the breaking wave kinematics and energy dissipation. The efficiency, flexibility, accuracy and stability of the model have been demonstrated for high-order Stokes wave propagation \cite{Wang_2022_AOR_FNPF}, irregular wave simulation \cite{Wang_2021_AOR_OE, Wang_2021_MDI}, focused wave generation \cite{Wang_JMSE_focus_2019} and wave transformations in the coastal waters and harbours \cite{Wang_2022_AOR_FNPF, Wang_2022_AOR_FNPF}.

Further advancements in numerical modeling have integrated wave-structure interaction (WSI) analysis within phase-resolving frameworks. For instance, the Arbitrary Lagrangian-Eulerian (ALE) method in REEF3D::FNPF enables real-time force estimations on structures without requiring explicit geometric modeling \citep{Pakozdi_2022_ALE_MS}. This method, utilizing nonlinear wave kinematics and the Morison equation \cite{morison1954}, facilitates rapid force estimations. With calibrations in mind, the method can be adapted to include slamming loads as well \citep{Pakozdi_2022_ALE_Slamming, Kamath_2023_ALE_Slamming}. Though the detailed wave-structure interaction is not resolved, the method can provide force estimation for multiple structures in the efficient wave model REEF3D::FNPF at runtime and presents an opportunity for wind-farm-scale preliminary designs. The irregular wave simulations often require over 3 hours of duration for a sufficient statistical representation. Consequently, the ALE approach is able to present 3-hour time series of the wave forces which act as an efficient extreme events identification (EEI) screening. 

For extreme limit state designs that require breaking-wave-structure interaction analysis, viscous Navier-Stokes solvers with turbulence models can be employed. These computational fluid dynamics (CFD) models effectively capture complex free surface interactions, including overturning plunging breakers, providing a more comprehensive and detailed slamming load calculation than the calibrated ALE approach. However, these models demand significant computational resources and time. A common strategy to enhance efficiency while maintaining flow detail resolution is coupling a non-viscous solver with a CFD model. For instance, Technip has developed a hydrodynamic coupling (HDC) between its in-house potential flow solver TPNWT and the commercial CFD software StarCCM+ for sea-state statistical analysis \citep{baquet2017omae}. Similarly, the potential flow solver OceanWave3D \citep{EngsigKarup_2009} has been integrated with the numerical wave tank (NWT) in the CFD solver OpenFOAM \citep{jacobsen2012} to analyze breaking wave forces on a vertical cylinder \citep{paulsen2014}. The open-source hydrodynamic framework REEF3D offers multiple models and built-in HDC protocols \citep{Wang_2022_jomae}, facilitating a more seamless coupling within the same numerical framework while utilizing shared parallelized high-performance computing (HPC) capabilities. The HDC protocol is explained in detail through the example of coupling the fully nonlinear potential flow model REEF3D::FNPF \citep{Wang_2022_AOR_FNPF} with the CFD solver REEF3D::CFD \citep{bihs2016cf}. However, the modular design of the REEF3D framework enables a straightforward extension of the coupling protocol to the non-hydrostatic shallow water equation-based model REEF3D::SFLOW \citep{Wang_2020_IJNMF} as demonstrated in \citep{Dempwolff_2023_SFLOW_HDC}. The CFD model REEF3D::CFD has undergone extensive validation for various breaking wave interactions with cylindrical and jacket structures \citep{alagan2019jfs, Aggarwal2019jfs, AGGARWAL2020108098}. The HDC procedures effectively link far-field wave kinematics with near-field fluid-structure interaction (FSI) analysis under breaking waves within this framework. However, the previous HDC investigations tend to focus on controlled lab environments with relatively simple bathymetry such as a submerged slope. The HDC method's applicability and practicality in real-world scenarios with natural bathymetry need further demonstrations. 

The development of open-access databases, open-source models, and HDC protocols paves the way for a structured approach to downscaling large-scale offshore met-ocean data for near-field FSI analysis of breaking waves. Advancing beyond conventional approaches, this study proposes a streamlined multi-scale synergy that effectively integrates individual models and techniques to establish an efficient connection between metocean data and FSI analysis. The proposed framework is codenamed NORA-SARAH (\textbf{NORA}-\textbf{S}W\textbf{A}N-\textbf{R}EEF3D-\textbf{A}LE-\textbf{H}DC) which combines the NORA metocean databases, the SWAN phase-averaged wave model, the REEF3D::FNPF phase-resolving wave model, the ALE force estimation method, and the HDC protocol within the REEF3D modeling environment. While each modeling component has been previously investigated, this framework uniquely integrates open-access data and open-source models to facilitate realistic, time-efficient, and adaptable hydrodynamic analyses for coastal and marine structures. The resulting near-field CFD simulation provides high-resolution insights, demonstrating the feasibility of the HDC protocol in accurately capturing complex wave dynamics over natural bathymetry within a large-scale engineering application. Furthermore, the framework achieves this level of detail while optimizing computational efficiency. A case study in southern Norway demonstrates the workflow and effectiveness of the proposed approach. The findings underscore the necessity of adopting the NORA-SARAH framework, or similar multi-scale methodologies, over conventional spectral wave models or simplified design approaches based on significant wave height ($H_s$).

\section{Numerical models}
\label{sec2:numerics}
\subsection{Phase-averaging wave model}

The SWAN (Simulating Waves Nearshore) model \citep{Booij_1999} is a third-generation spectral wave model designed to simulate wave energy distribution and transformation. It achieves this by solving the spectral action balance equation, which can be expressed in a Cartesian coordinate system as:

\begin{equation}
\label{eq_swan}
\frac\partial{\partial t}N+\frac\partial{\partial x}c_xN+\frac\partial{\partial y}c_yN+\frac\partial{\partial\sigma}c_\sigma N+\frac\partial{\partial\theta}c_\theta N=\frac{S_{tot}}{\sigma},
\end{equation}

where $N$ represents the spectral action density, defined as $E/\sigma$, with $E$ being the spectral energy density and $\sigma$ the relative wave frequency in the moving frame of the current. The propagation velocities $c_x$, $c_y$, $c_\sigma$, and $c_\theta$ govern the spatial, frequency, and directional evolution of the spectral action. The term $S_{tot}$ represents the total source term, encompassing the various physical processes influencing wave transformation:

\begin{equation}
\label{eq_source}
S_{tot}=S_{in}+S_{nl3}+S_{nl4}+S_{ds,w}+S_{ds,b}+S_{ds,br}.
\end{equation}

Here, $S_{in}$ accounts for wind energy input, $S_{nl3}$ and $S_{nl4}$ model triad and quadruplet nonlinear wave-wave interactions, while $S_{ds,w}$, $S_{ds,b}$, and $S_{ds,br}$ represent dissipation mechanisms due to white-capping, bottom friction, and wave breaking in shallow water, respectively.

\subsection{Phase-resolving models}

The open-source hydrodynamic code REEF3D \citep{bihs2016cf} offers various phase-resolving numerical models, each tailored for different scales and physical phenomena. For large-scale coastal wave propagation, the fully nonlinear potential flow model REEF3D::FNPF \citep{Wang_2022_AOR_FNPF} is employed, while near-field wave-structure interactions are captured using the computational fluid dynamics (CFD) solver REEF3D::CFD \citep{bihs2016cf}.

\subsubsection{REEF3D::FNPF}

REEF3D::FNPF is based on the solution of the Laplace equation for the velocity potential $\phi$:

\begin{align}
\frac{\partial^2\phi}{\partial x^2}+\frac{\displaystyle\partial^2\phi}{\displaystyle\partial y^2}+\frac{\displaystyle\partial^2\phi}{\displaystyle\partial z^2}=0. \label{eq:laplace}
\end{align}

Boundary conditions include kinematic and dynamic free surface conditions and the kinematic bottom boundary condition:

\begin{align}
\frac{\partial\phi}{\partial z}+\frac{\partial h}{\partial x}\frac{\partial\phi}{\partial x}+\frac{\partial h}{\partial y}\frac{\partial\phi}{\partial y}=0, \;\;z=-h, \label{eq:bot_bc}
\end{align}

\begin{align}
\frac{\partial\eta}{\partial t}=-\frac{\partial\eta}{\partial x}\frac{\partial\widetilde\phi}{\partial x}-\frac{\partial\eta}{\partial y}\frac{\partial\widetilde\phi}{\partial y}
+\widetilde w\left(1+\left(\frac{\partial\eta}{\partial x}\right)^2+\left(\frac{\partial\eta}{\partial y}\right)^2\right),  \label{eq:kf_bc} 
\end{align}

\begin{align}
\frac{\partial\widetilde\phi}{\partial t}=-\frac12\left(\left(\frac{\displaystyle\partial\widetilde\phi}{\displaystyle\partial x}\right)^2+\left(\frac{\displaystyle\partial\widetilde\phi}{\displaystyle\partial y}\right)^2\right) 
+\frac12\widetilde w^2\left(1+\left(\frac{\displaystyle\partial\eta}{\displaystyle\partial x}\right)^2+\left(\frac{\displaystyle\partial\eta}{\displaystyle\partial y}\right)^2\right)-g\eta,  \label{eq:df_bc} 
\end{align}

where the velocity potential and the vertical velocity at the free surface $\eta$ are identified with a tilde $\widetilde{\phi}$ and $\widetilde{w}$ and the horizontal coordinate is written as $\textbf{x}=(x,y)$. \\

The model employs a finite difference approach and utilizes the BiCGStab solver \cite{vorst1992} preconditioned with the geometric multigrid solver PFMG \cite{ashby1996} from the hypre library. While the horizontal grid remains rectilinear, a $\sigma$-grid is implemented in the vertical direction, ensuring increased resolution near the free surface:

\begin{equation}
\sigma=\frac{z+h\left(\textbf{x}\right)}{\eta(\textbf{x},t)+h(\textbf{x})},
\label{eq_sigma}
\end{equation}

where $h(\textbf{x})$ is the local wave depth in the Cartesian coordinate. 

Wave breaking is identified using depth- and steepness-based criteria \cite{zijlema_2011, Smit_2013}, and energy dissipation is introduced through an artificial viscous term \cite{Baquet_OMAE_2017}. 

Coastlines are captured using a level-set function, allowing a smooth transition between wet and dry cells, which are designated with opposite signs and identified based on a local water depth threshold.

\subsubsection{Arbitrary Eulerian-Lagrangian (ALE) force calculation}

The $\sigma$-grid formulation provides precise free surface locations, enabling the representation of nonlinear wave kinematics in an Arbitrary Eulerian-Lagrangian (ALE) framework \citep{Donea_2004_ale, Pakozdi_2022_ALE_MS}. Following the ALE approach, the fluid particle acceleration is written as:

\begin{align}
\label{eq::axALE}
a_x  =  \left. \frac{\partial u}{\partial t}\right|_{\vec{\sigma}} + u\left( \frac{\partial u}{\partial \xi}+\dfrac{\partial u}{\partial \sigma}\dfrac{\partial \sigma}{\partial x} \right) 
 +\left(w-\sigma \left.\dfrac{\partial \eta(x,t)}{\partial t} \right|_{\vec{x}} \right) \frac{\partial u}{\partial \sigma}\frac{\partial \sigma}{\partial z},
\end{align}

where the horizontal and vertical velocities and coordinates are $u$, $w$, $\xi$ and $\sigma$ respectively. 

The nonlinear wave velocity and accelerations absence from the disturbance from the structures integrated the Morison equation for rapid wave forces calculations: 

\begin{align}
\label{eq::globalforcesALE}
	F_x =\rho (h+\eta) 
	 \left[ \int_0^{1}C_M a_x A_{xy}  d\sigma+\int_0^{1}C_D u|u| \dfrac{1}{2} B_p d\sigma \right],
\end{align}

where $C_M$ and $C_D$ are inertia and drag coefficients, while $A_{xy}$ and $B_p$ denote the cross-section area and section width.

With calibrations, the adapted form of the ALE method can even account for slamming loads as well\citep{Pakozdi_2022_ALE_Slamming, Kamath_2023_ALE_Slamming}. 

\subsubsection{REEF3D::CFD}

REEF3D::CFD \cite{bihs2016cf} employs a finite difference method to solve the Navier-Stokes equations:

\begin{equation}\label{eq:con}
\frac{\partial {u_i}}{\partial x_{i}}=0,
\end{equation}

\begin{equation}\label{eq:rans}
\frac{\partial u_i}{\partial t}+ u_j\frac{\partial u_i}{\partial x_j}=-\frac{1}{\rho}\frac{\partial p}{\partial x_i}+\frac{\partial}{\partial x_j}\left[(\nu+\nu_t)\left(\frac{\partial u_i}{\partial x_j}+\frac{\partial u_j}{\partial x_i}\right)\right]+ g_i,
\end{equation}

where $u$ represents the velocity of the fluid particles, $\rho$ is the fluid density, $p$ is pressure, $\nu$ denotes the kinematic viscosity, $\nu_t$ is the eddy viscosity, and $g$ accounts for gravitational acceleration.

The pressure solution is obtained using Chorin's projection method \cite{chorin1968}, and the Poisson equation is handled via the BiCGStab solver \cite{vorst1992}, preconditioned with the geometric multigrid PFMG from the hypre library for enhanced computational efficiency.

This two-phase model resolves both water and air domains. The free surface, representing the interface between these two phases, is captured through the level-set function \cite{osher1988}, which is defined as a signed distance function:

\begin{equation}\label{levelset1}
\phi_l(\vec{x},t)\begin{cases} >0\ if\  \vec{x}\in phase \ 1, \\
0\ if\  \vec{x}\in \Gamma,\\
<0\ if\  \vec{x}\in phase \ 2. \\
\end{cases} \end{equation}

The distance function property is maintained by ensuring the validity of the Eikonal equation $\left|\nabla \phi_l \right| = 1$.

REEF3D::CFD incorporates a range of turbulence models, including Reynolds-averaged Navier-Stokes (RANS), large-eddy simulations (LES), and direct numerical simulations (DNS). For the present study on breaking wave interactions, the RANS approach with the two-equation $k$-$\omega$ model \cite{wilcox1994} is applied.

Both REEF3D models implement high-order numerical schemes for improved accuracy. The fifth-order WENO (weighted essentially non-oscillatory) scheme \cite{jiang1996} is used for spatial discretization, while the third-order Runge-Kutta scheme \cite{shu1988} is adopted for temporal advancement. These high-order schemes are particularly beneficial for accurately capturing steep and nonlinear wave characteristics.

The two phase-resolving models also share wave generation methods. Surface waves can be generated following the relaxation zone method \cite{larsen1983, mayer1998} where one characteristic wavelength is typically used as the relaxation zone length \cite{bihs2016cf}. To reduce the cell counts, boundary conditions prescribing wave kinematics and surface elevations can be used for wave generation as well, specifically a Dirichlet boundary condition for the CFD model and a Neumann boundary condition for the FNPF model. To mitigate undesired reflections from the outlet boundaries, a relaxation zone is often used as a numerical beach following a reverse process of the wave generation relaxation zone. Alternatively, active absorptions can be utilized to cancel the incident waves following the shallow water linear wave theory. 

Furthermore, all REEF3D models are fully parallelized using domain decomposition, with subdomains communicating through the Message Passing Interface (MPI) protocol. This enables highly scalable simulations, allowing for efficient utilization of large numbers of processors to significantly enhance computational speed.

\subsubsection{Hydrodynamic Coupling (HDC)}

Within the REEF3D framework, a one-way coupling strategy facilitates the transfer of simulation data between different models \cite{Wang_2021_HDC, Dempwolff_2023_SFLOW_HDC}. During the FNPF simulation, the hydrodynamic state at each time step or iteration is stored in dedicated state files covering the entire computational domain. Once the FNPF simulation is completed, these state files can be processed and interpolated onto the CFD numerical wave tank (NWT) computational grid. The interface boundary between the domains can be performed either following a relaxation zone strategy or by a Dirichelt boundary. The CFD simulation can then be performed using the stored flow data, employing either the same number of processors as the FNPF run or an entirely different high-performance computing (HPC) configuration. 

This approach enhances computational flexibility, allowing multiple CFD simulations to be initiated from different time points and spatial locations while leveraging data from a single FNPF simulation. As a result, overall computational efficiency is improved by maximizing data reusability and minimizing redundant simulations. The synergy between these models ensures a streamlined workflow for detailed hydrodynamic analysis of breaking wave interactions.

\section{The NORA-SARAH Framework}

\begin{figure}[!hptb]
\centering
\includegraphics[width=0.5\textwidth]{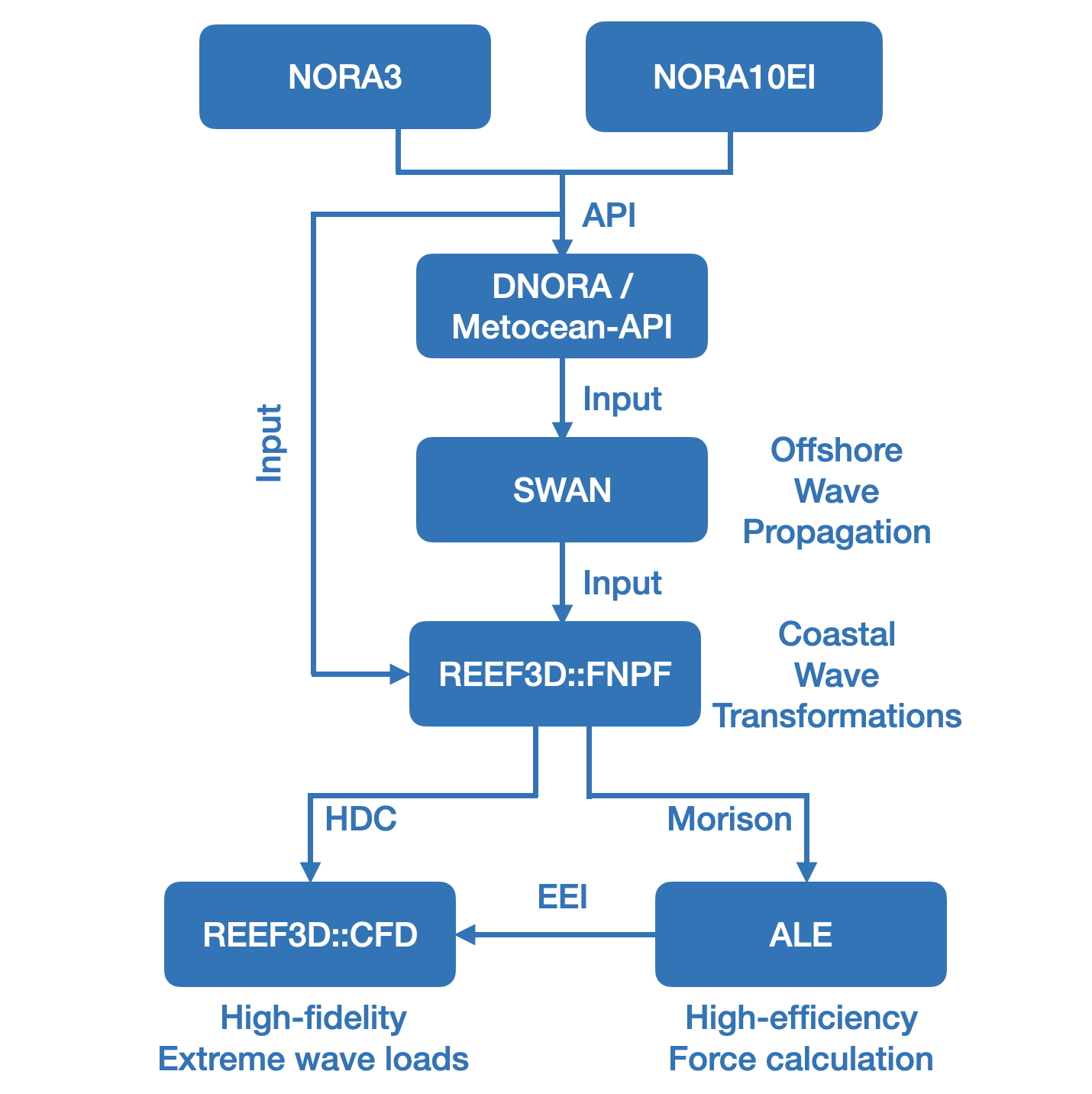}
\caption{Schematic representation of the NORA-SARAH framework (\underline{NORA}-\underline{S}W\underline{A}N-\underline{R}EEF3D-\underline{A}LE-\underline{H}DC), illustrating the integration of offshore hindcast data with numerical modeling for marine structure wave load estimation.}
\label{fig::nora-sarah}
\end{figure}

The open-access met-ocean databases NORA10EI and NORA3 serve as essential sources of offshore environmental data, which can be systematically downscaled for near-field wave-structure interaction analysis. This information propagation is facilitated by a sequence of open-source numerical models. The open-source metocean-api (https://github.com/MET-OM/metocean-api) provided by MET is used to extract met-ocean conditions from the NORA3 wind and wave database for a specified location and time window, converting them into compatible inputs for wave models such as SWAN and REEF3D. The phase-averaging model SWAN provides statistical wave characteristics and bridge offshore and nearshore wave conditions. In contrast, the phase-resolving REEF3D::FNPF model captures nonlinear wave transformations influenced by seabed variations and coastal features. For wave load assessment, the ALE method operates in real-time within REEF3D::FNPF simulations, estimating wave-induced forces for preliminary design and extreme event identification (EEI). The wave kinematics computed within the FNPF numerical wave tank (NWT) are then mapped onto the computational domain of the viscous CFD solver REEF3D::CFD, enabling detailed analysis of breaking-wave-induced forces under extreme conditions. This integrated workflow, referred to as the NORA-SARAH approach (\textbf{NORA}-\textbf{S}W\textbf{A}N-\textbf{R}EEF3D-\textbf{A}LE-\textbf{H}DC), structures the transition from large-scale offshore hindcast data to high-fidelity numerical modeling for wave load estimation, as illustrated in Fig.~\ref{fig::nora-sarah}.

The subsequent section demonstrates the effectiveness of this framework through a case study on the southern Norwegian coast.

\section{Hydrodynamic analysis in Southern Norway}

\begin{figure}[!hptb]
\centering
\includegraphics[width=0.9\textwidth]{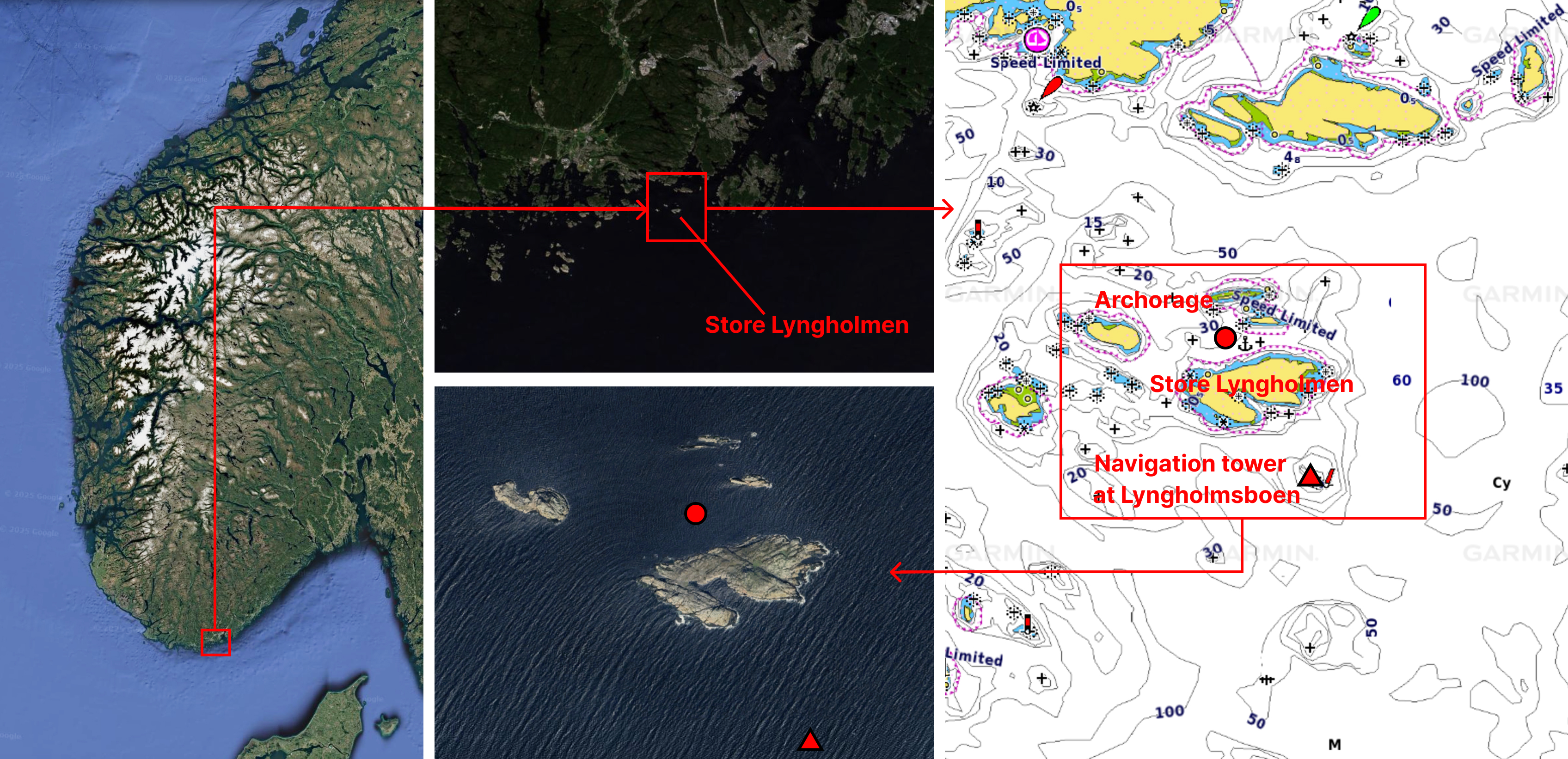}
\caption{Study site near the island of Store Lyngholmen in southern Norway. The left and center-upper panels display satellite images of the island and its surroundings, sourced from the Norwegian Mapping Authority (www.norgeskart.no). The right panel provides bathymetric data of the adjacent area, obtained from the Garmin Sea Chart (https://maps.garmin.com/nb-NO/marine). The center-lower panel presents a satellite snapshot from Google Earth (https://earth.google.com/web), capturing wave propagation and diffraction around the island. The navigation tower location is marked with a red triangle, and the anchorage site is indicated with a red circle.}
\label{fig::study-site}
\end{figure}

The selected study site for demonstrating the NORA-SARAH framework is the water body surrounding the island of Store Lyngholmen, located near Kristiansand in southern Norway. The location, along with its surrounding topographic and bathymetric environment, is illustrated in Fig.~\ref{fig::study-site}. Situated at Norway’s southern tip, the site is exposed to southern swell and wind waves originating from both the Atlantic Ocean to the west and the Skagerrak Strait to the east, which serves as a connection between the Atlantic Ocean and the Baltic Sea.

A navigation tower, marked as a red triangle in Fig.\ref{fig::study-site}, is proposed to be constructed on a shoal known as Lyngholmsboen (58.052004$^{\circ}$ N, 7.925912$^{\circ}$ E), located southeast of Store Lyngholmen. This shoal features a peak water depth of approximately 5 m, while the surrounding waters reach depths of roughly 50 m. Constructing the tower in this shallow-water region offers advantages such as reduced material requirements and lower construction costs. However, the presence of the shoal also influences wave behavior, leading to wave shoaling, refraction, and an increase in wave energy concentration. These effects can introduce nonlinear wave kinematics and higher-order wave harmonics. The horizontal distance from the shoal’s peak to the 50 m depth contour to the south is approximately 185 m, resulting in a submarine slope of about 1:4. This relatively steep slope poses challenges related to wave breaking and potential slamming loads on the structure. The NORA-SARAH framework will be employed to analyze wave conditions and wave-induced loads, including slamming effects on the tower.

On the leeward side of Store Lyngholmen, an anchorage site (58.0579741$^{\circ}$ N, 7.920036$^{\circ}$ E) is marked as a red circle in Fig.\ref{fig::study-site}. The satellite image in the center-lower panel of Fig.\ref{fig::study-site} shows reduced wave activity in the diffraction zone on the northern side of the island. However, varying wave conditions can result in different diffraction intensities and changes in wave energy distribution across frequencies due to the non-uniform nature of the diffracted wave field. As discussed in the introduction, traditional spectral wave models often fail to capture diffraction effects in sufficient detail. Consequently, wave conditions at the anchorage site will also be analyzed using the NORA-SARAH framework.

\begin{figure}[!hptb]
 	\centering
 	\includegraphics[width=0.99\textwidth]{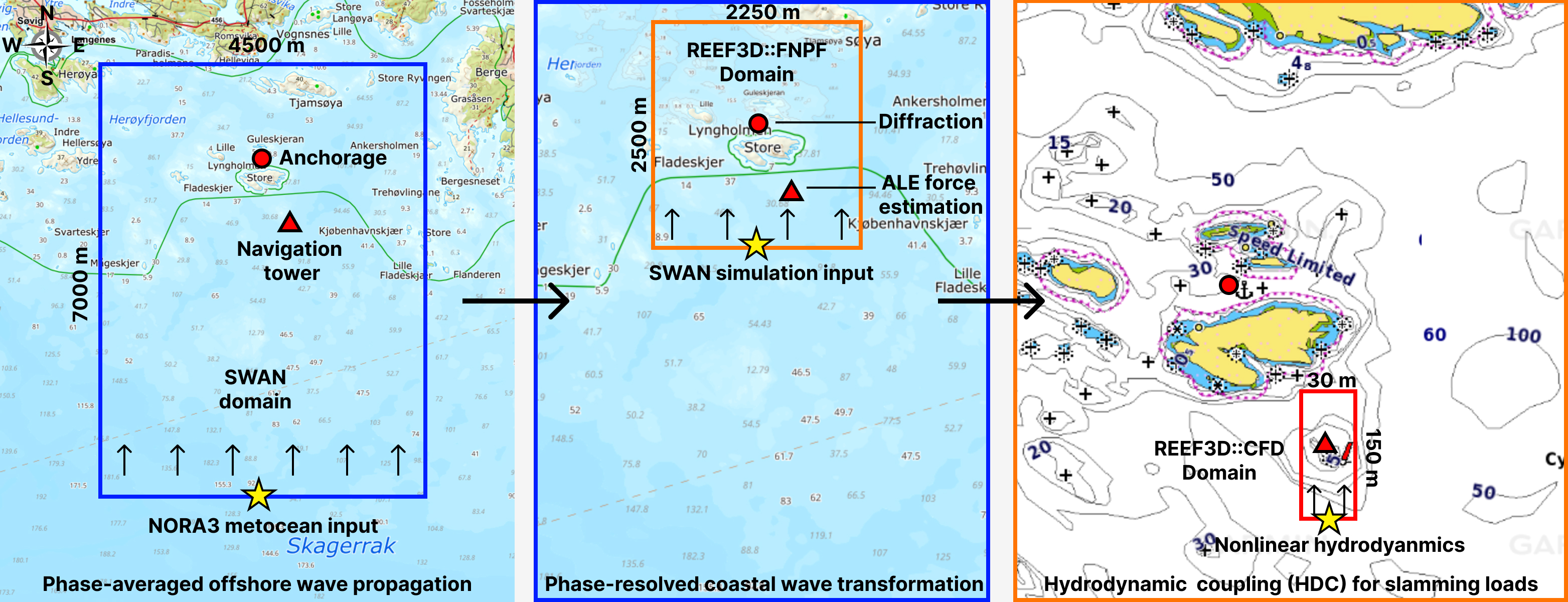}
 	\caption{The down-scale process with reduced domain sizes from the phase-averaged offshore wave propagation simulation with SWAN (marked as a blue box) to the phase-resolved coastal wave transformation simulation with REEF3D::FNPF (marked as an orange box) further to the near-field fluid-structure interaction simulation using REEF3D::CFD (marked as a red box) follow an HDC protocol with Dirichlet coupling boundary conditions. The yellow stars represent the wave input between the scales.} 
 	\label{fig:down-scale}
\end{figure}

\begin{figure}[!hptb]
 	\centering
 	 	\includegraphics[width=0.8\textwidth]{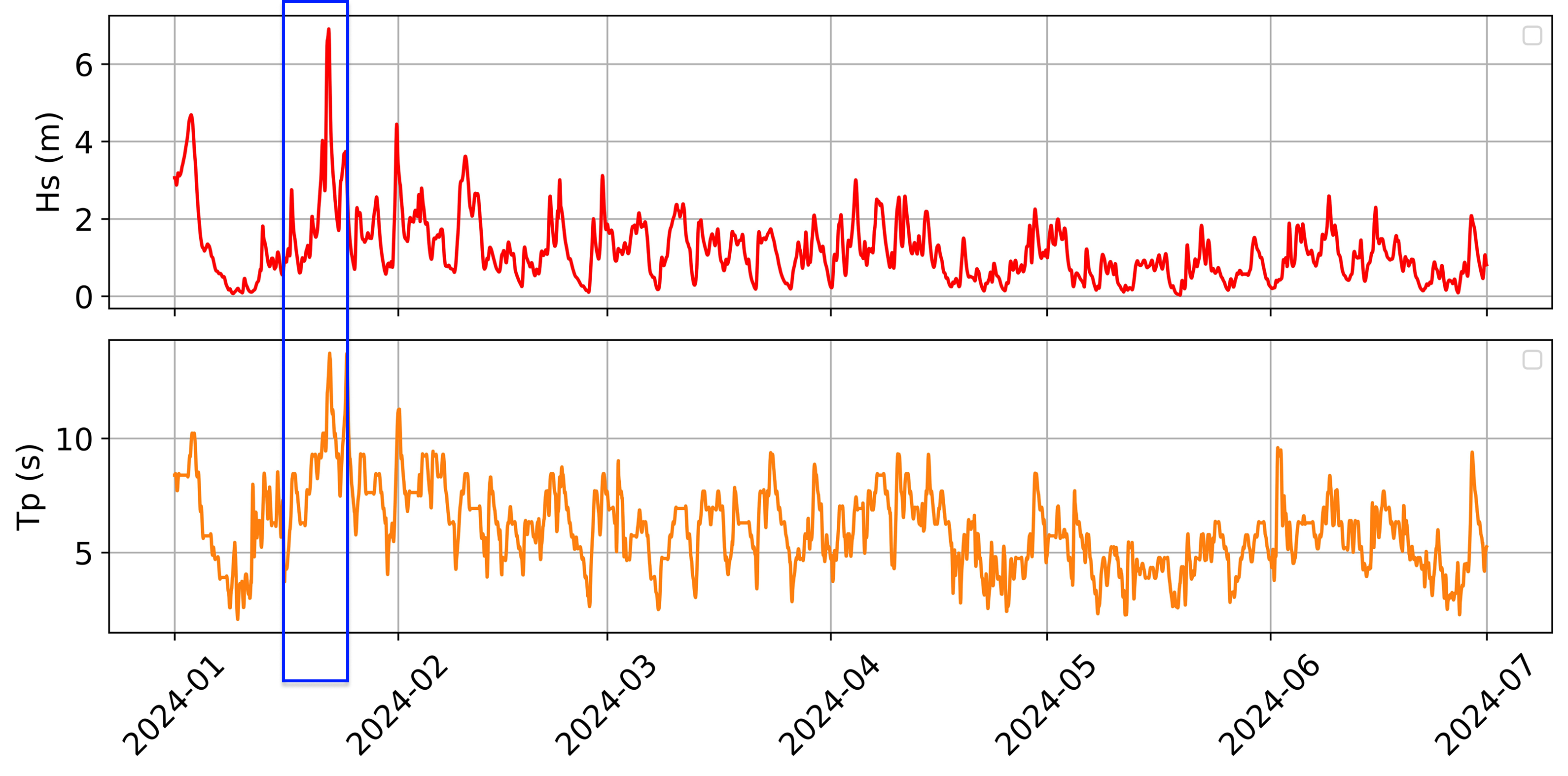}
 	\caption{The offshore $H_s$ and $T_p$ time series between 2024-01-01 and 2024-06-30. } 
 	\label{fig:metocean}
\end{figure}

The downscaling analysis following the NORA-SARAH strategy is summarized and visualized in Fig.\ref{fig:down-scale}. Metocean data is obtained from an offshore location (58.00711$^{\circ}$ N, 7.93309$^{\circ}$ E), marked as a yellow star in the left panel of Fig.\ref{fig:down-scale}, using metocean-api linked to the NORA3 wind and wave hindcast database. This location is situated near the 100 m depth contour, representing intermediate to deep water conditions for most offshore irregular wave components. The site is approximately 4 km from the Lyngholmsboen shoal, exceeding the 3 km resolution of NORA3, ensuring adequate metocean data interpolation. The local significant wave height and peak period information between 2024-01-01 and 2024-06-30 are obtained and summarised in Fig.~\ref{fig:metocean}. For the purpose of demonstration, long-crested waves coming from  $180^{\circ}$ south are assumed during the period despite the actual temporal variations. It is seen that the most severe sea state is observed in the winter season, especially during late January, where a maximum significant wave height ($H_s$) and peak period ($T_p$) combination is recorded as highlighted in the blue box in Fig.~\ref{fig:metocean}. The maximum $H_s$ and $T_p$ are 6.88 m and 13.67 s, which are used as input parameters for the offshore phase-averaged SWAN simulations. 

The SWAN simulation domain, outlined as a blue box in Fig.~\ref{fig:down-scale}, extends 7 km northward from the metocean input location at its southern boundary, with an east-west span of 4.5 km. This domain encompasses key topo-bathymetric features that significantly influence wave propagation and transformation toward the site of interest. To ensure uniform offshore wave generation and minimize diffraction effects near the generation boundary, wave input is applied at three boundaries: south, west, and east. The imposed long-crested wave field at $180^{\circ}$ is approximated in SWAN by setting the power parameter $m$ to 130 in the $cos^m(\theta - \theta_{peak})$ directional spreading function, corresponding to a $5^{\circ}$ standard deviation in directional spreading.

A horizontal resolution of 10 m is used in the SWAN simulations. The directional sector from $90^{\circ}$ to $270^{\circ}$ is discretized into 72 bins, each with an angular resolution of $2.5^{\circ}$. The frequency range spans from half the peak frequency to three times the peak frequency, resulting in a range of [0.035, 0.215] Hz, discretized with 0.001 Hz intervals. The wave spectrum follows a JONSWAP distribution, an appropriate assumption given the site's exposure to the North Atlantic Ocean. The simulation is conducted as a stationary case without additional wind forcing.

\begin{figure}[!hptb]
 	\centering
 	 	\includegraphics[width=0.5\textwidth]{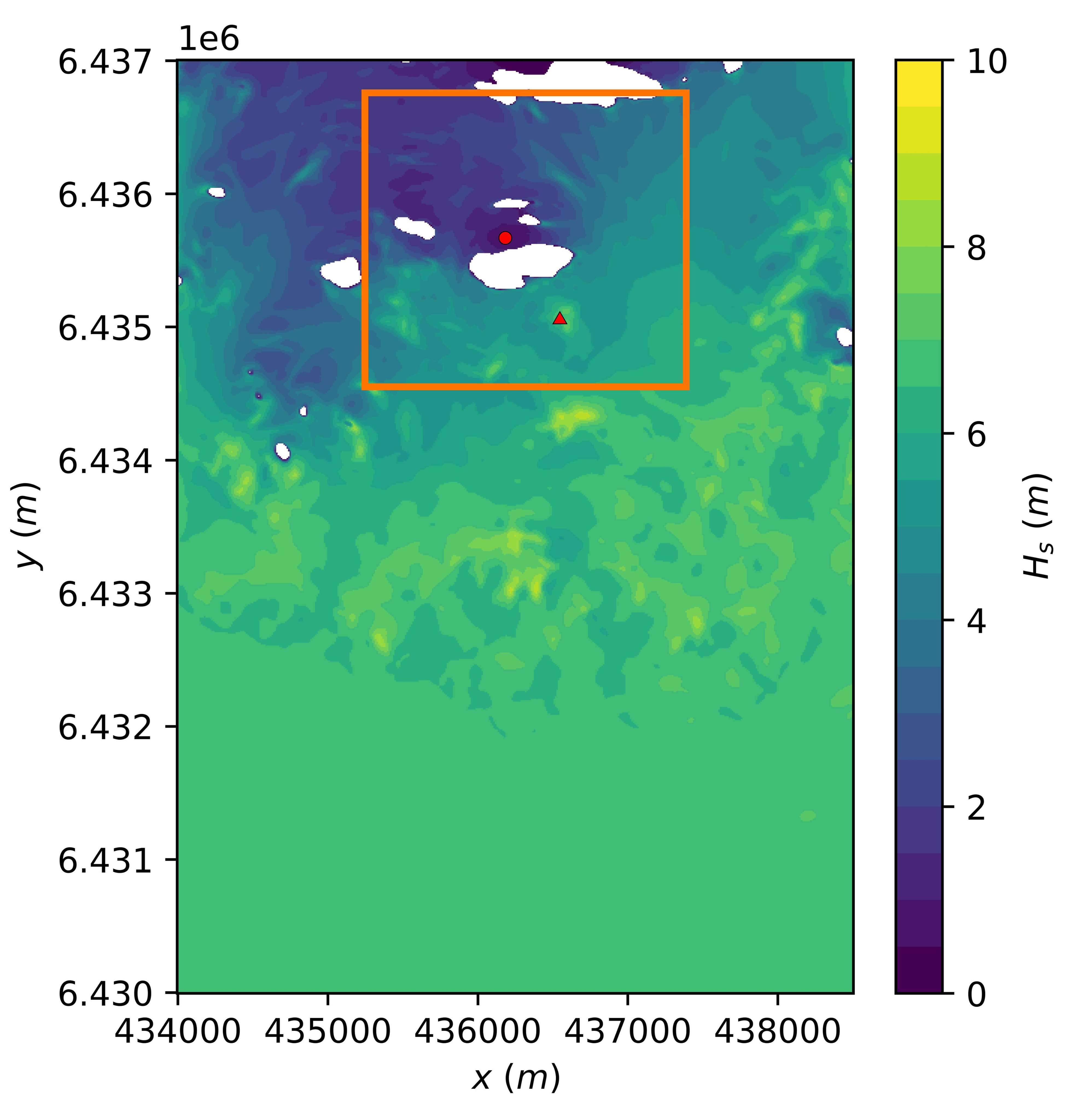}
 	\caption{$H_s$ distribution from the SWAN simulation with the NORA3 inputs. The orange box shows the REEF3D phase-resolving simulation domain within the SWAN domain.} 
 	\label{fig::SWAN}
 \end{figure}

The spatial distribution of significant wave height ($H_s$) within the simulation domain is illustrated in Fig.\ref{fig::SWAN}. While the offshore wave field remains homogeneous, bathymetry-induced wave transformations become evident in shallower waters. Notable increases in $H_s$ occur at multiple shoals are observed, including Lyngholmsboen, where the navigation tower is located. The SWAN output at the phase-resolving boundary, indicated by the yellow star in the center panel of Fig.\ref{fig:down-scale}, is used to define the input wave conditions for the REEF3D::FNPF numerical wave tank (NWT), represented by the orange box in Fig.~\ref{fig:down-scale}.

\begin{figure}[!hptb]
 	\centering
 	 	\includegraphics[width=0.5\textwidth]{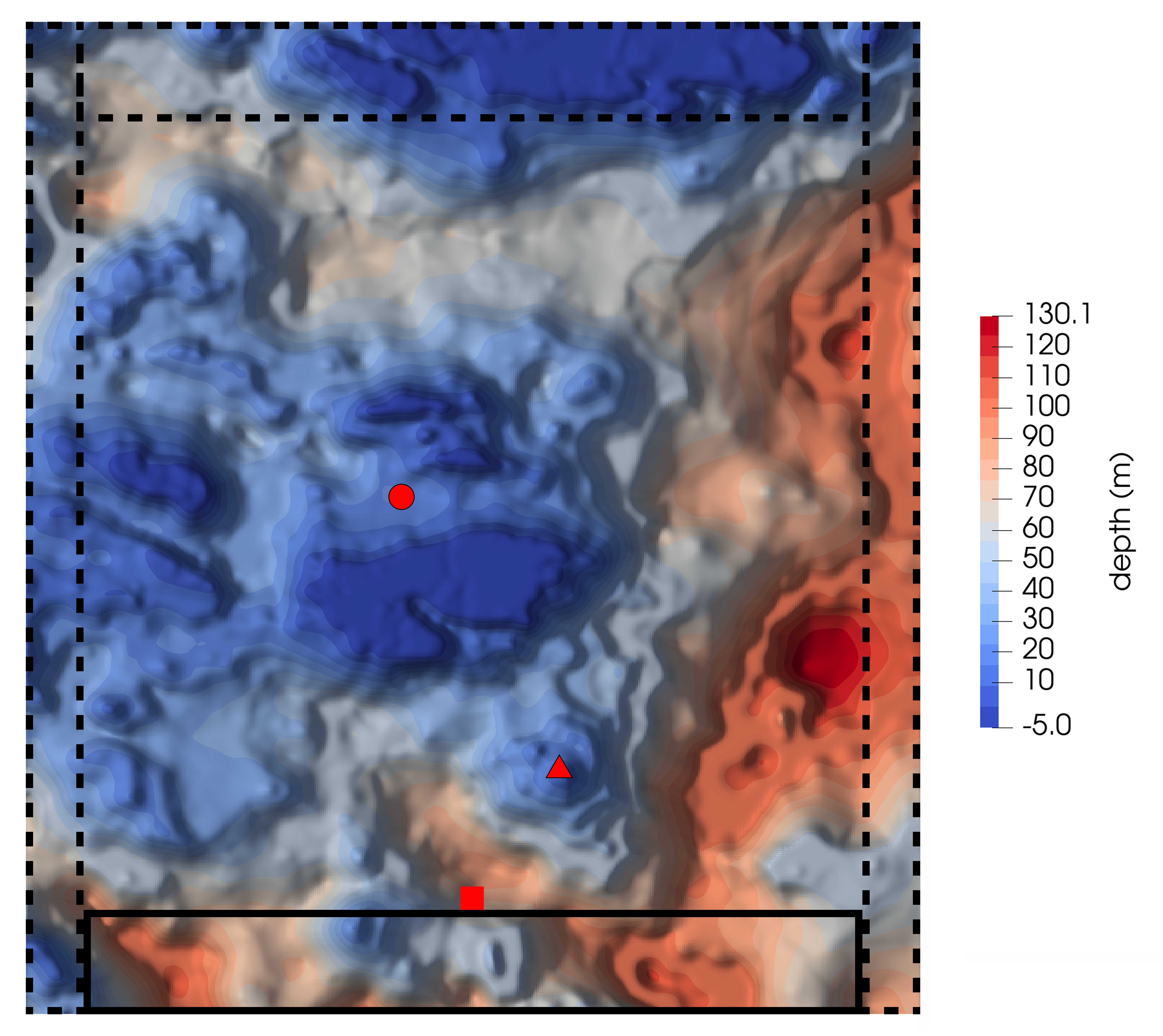}
 	\caption{The numerical representation of the bathymetry around Store Lyngholmen and the configuration of REEF3D::FNPF NWT.} 
 	\label{fig::fnfp_bathy}
 \end{figure}

The phase-resolving REEF3D::FNPF numerical wave tank (NWT) has a domain size of 2500 $\times$ 2250 m, which is less than one-quarter of the SWAN simulation domain. It focuses on the region surrounding Store Lyngholmen to provide detailed insights into wave transformations and estimate wave loads. The numerical representation of the local bathymetry and the NWT configuration are illustrated in Fig.~\ref{fig::fnfp_bathy}. The solid-line black box denotes the 400 m relaxation zone for wave generation, designed to accommodate two wavelengths corresponding to the peak period. The dashed-line black boxes indicate relaxation zones acting as numerical beaches to dissipate wave energy at the three remaining boundaries. The side boundaries feature 200 m relaxation zones, while the northern boundary along the principal wave propagation direction has a relaxation zone of 400 m.

Three wave gauges are deployed within the NWT, as shown in Fig.~\ref{fig::fnfp_bathy}. The red cube represents the wave gauge used to verify the input waves, the triangle indicates the wave gauge located at the navigation tower's center, and the circle marks the wave gauge positioned at the anchorage site. The domain is discretized with a uniform horizontal cell size of 5 m and 10 vertical cells, applying a stretching factor of 2.5 to refine resolution toward the free surface, resulting in a total of 2.25 million computational cells. Adaptive time stepping is implemented with a Courant–Friedrichs–Lewy (CFL) number of 1.0 to control time step sizes.

The input wave conditions are derived from the SWAN output and parameterized as a JONSWAP spectrum with $H_s$ = 6.0 m and $T_p$ = 12.0 s. The frequency range of [0.04, 0.25] Hz is discretized with 1024 frequency components. The simulation runs for 3.6 hours (12800 s), with the last 3 hours (10800 s) of free surface elevation data used for post-processing and wave spectrum reconstruction via Fast Fourier Transform (FFT). The computations are executed using 128 AMD EPYC 7763 cores on an Ubuntu-based workstation, achieving an elapsed runtime of 4.5 hours.

\begin{figure}[!hb]	
\centering
\begin{subfigure}[b]{0.45\textwidth}
                \centering
                \includegraphics[width=\textwidth]{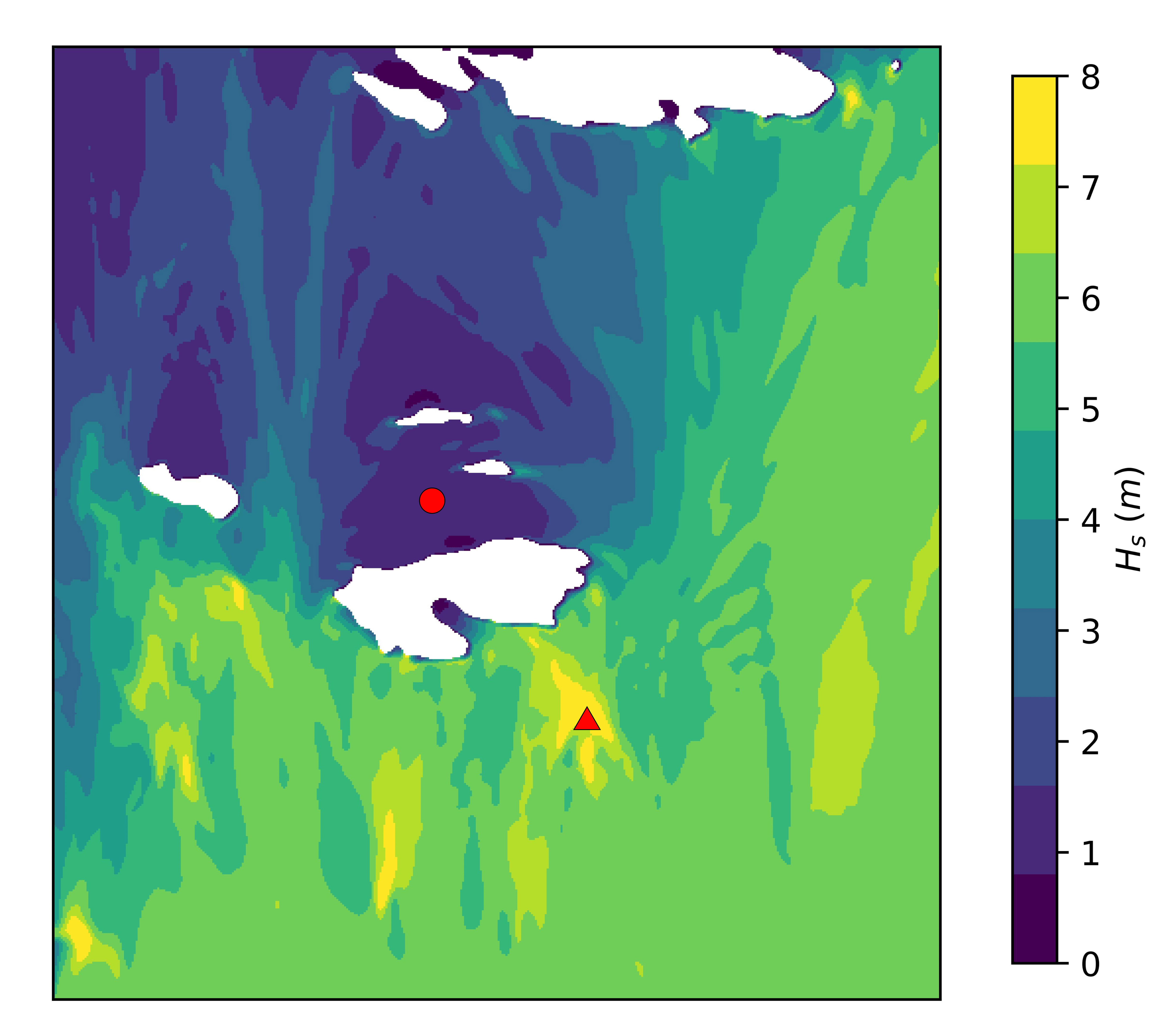}
                \caption{SWAN}
                \label{fig:swan_result}
\end{subfigure}
\begin{subfigure}[b]{0.45\textwidth}
                \centering
                \includegraphics[width=\textwidth]{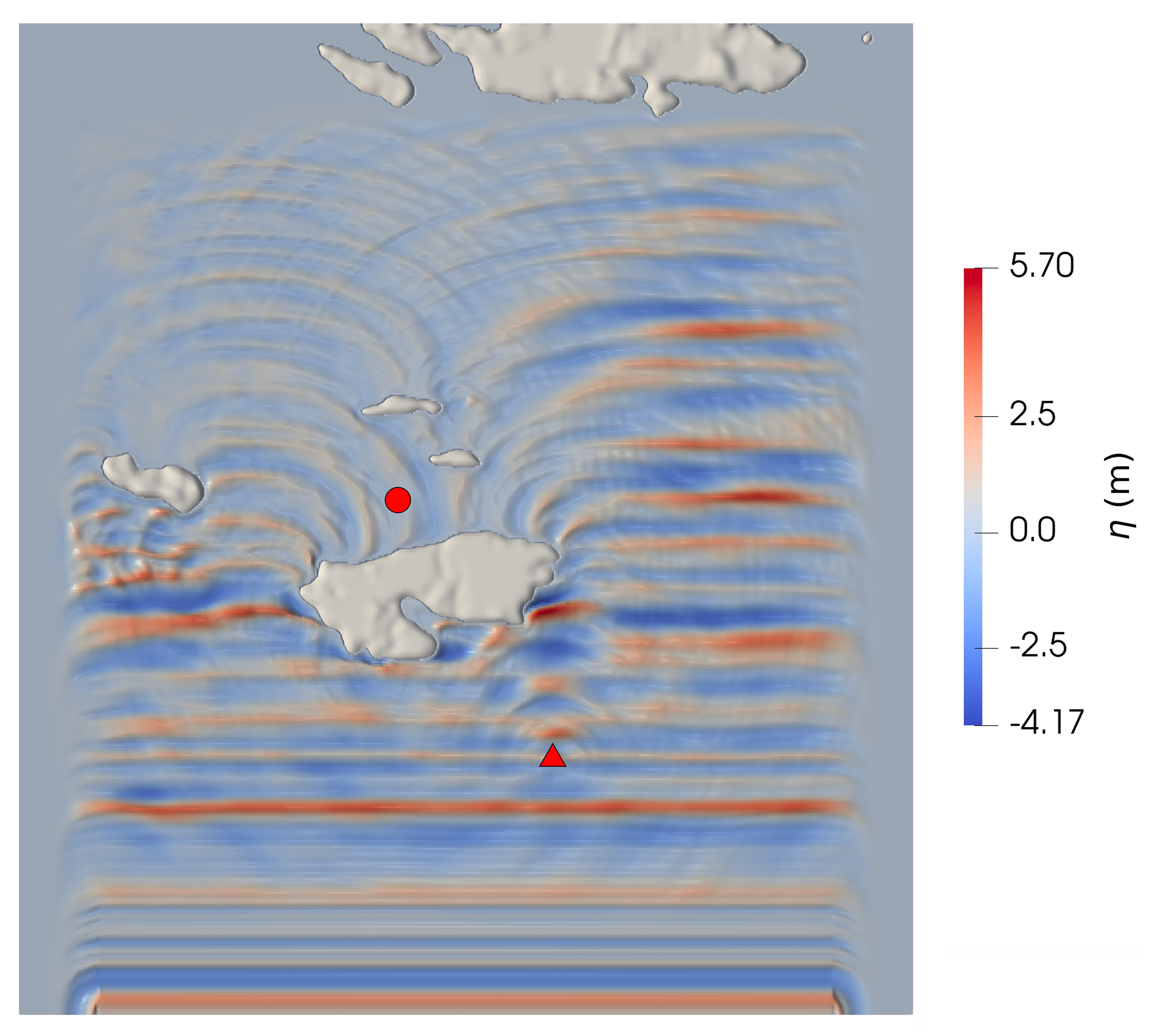}
                \caption{FNPF}
                \label{fig:fnpf_result}
\end{subfigure}
\caption{The $H_s$ distribution from the phase-averaging SWAN simulation and free surface elevation at 12800 s from the phase-resolving REEF3D::FNPF simulation.}
\label{fig:wave_result}
\end{figure}

To compare model representations of coastal wave transformations, a SWAN simulation is also performed using the same wave input, domain size, and horizontal grid resolution. The $H_s$ distribution from the SWAN simulation and the final free surface elevation at 12800 s from the REEF3D::FNPF simulation are shown in Fig.~\ref{fig:wave_result}. Both models capture the shoaling effects at Lyngholmsboen, but the diffraction patterns on the leeward side of Store Lyngholmen differ significantly. The SWAN simulation exhibits a more discrete energy distribution in the diffraction zone, whereas the phase-resolving approach in REEF3D::FNPF presents a more continuous redistribution of wave energy.

\begin{figure}[!hb]	
\centering
\begin{subfigure}[b]{0.3\textwidth}
                \centering
                \includegraphics[width=\textwidth]{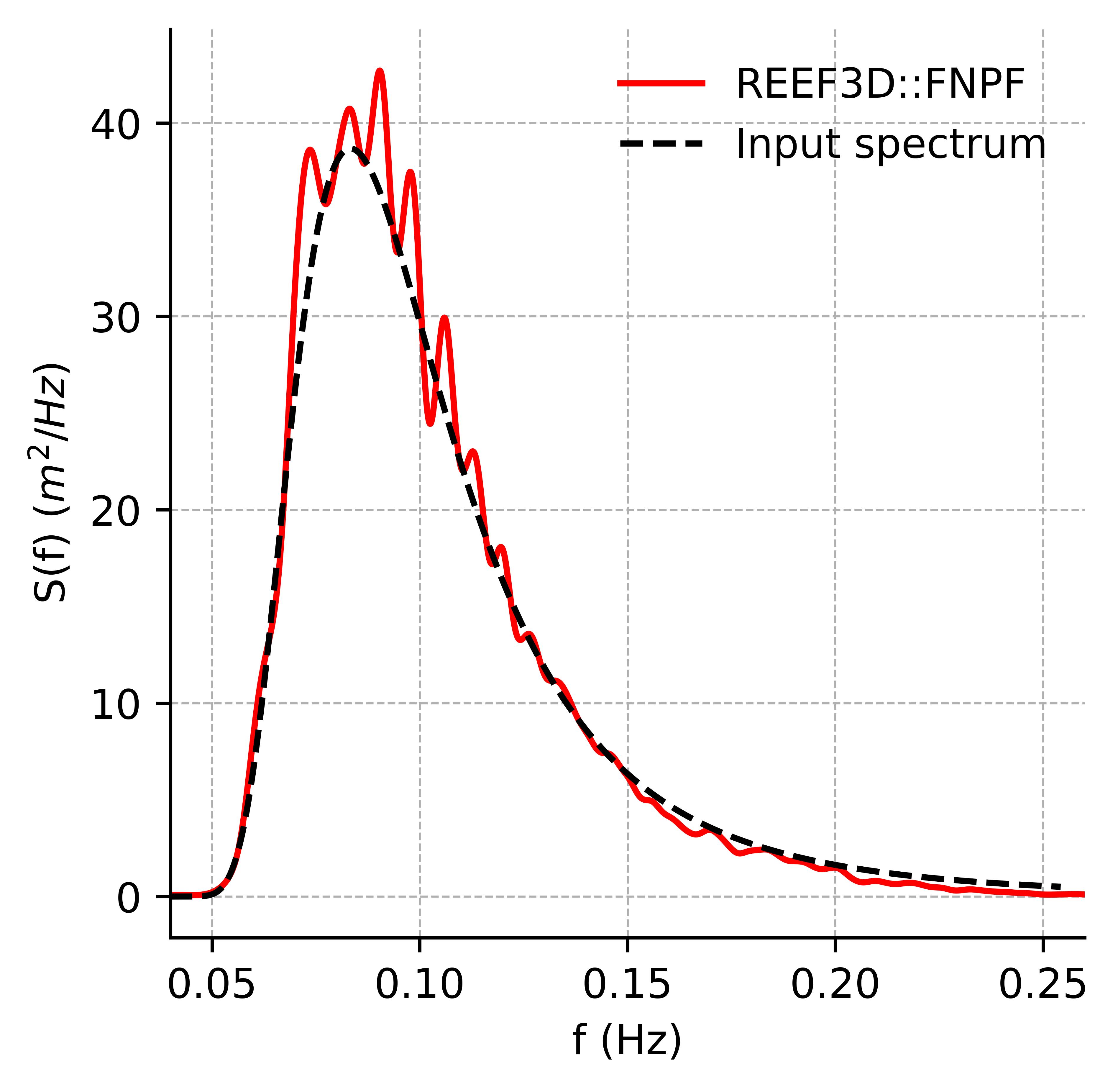}
                \caption{Verification wave gauge $\square$}
                \label{fig:gauge_1}
\end{subfigure}
\begin{subfigure}[b]{0.3\textwidth}
                \centering
                \includegraphics[width=\textwidth]{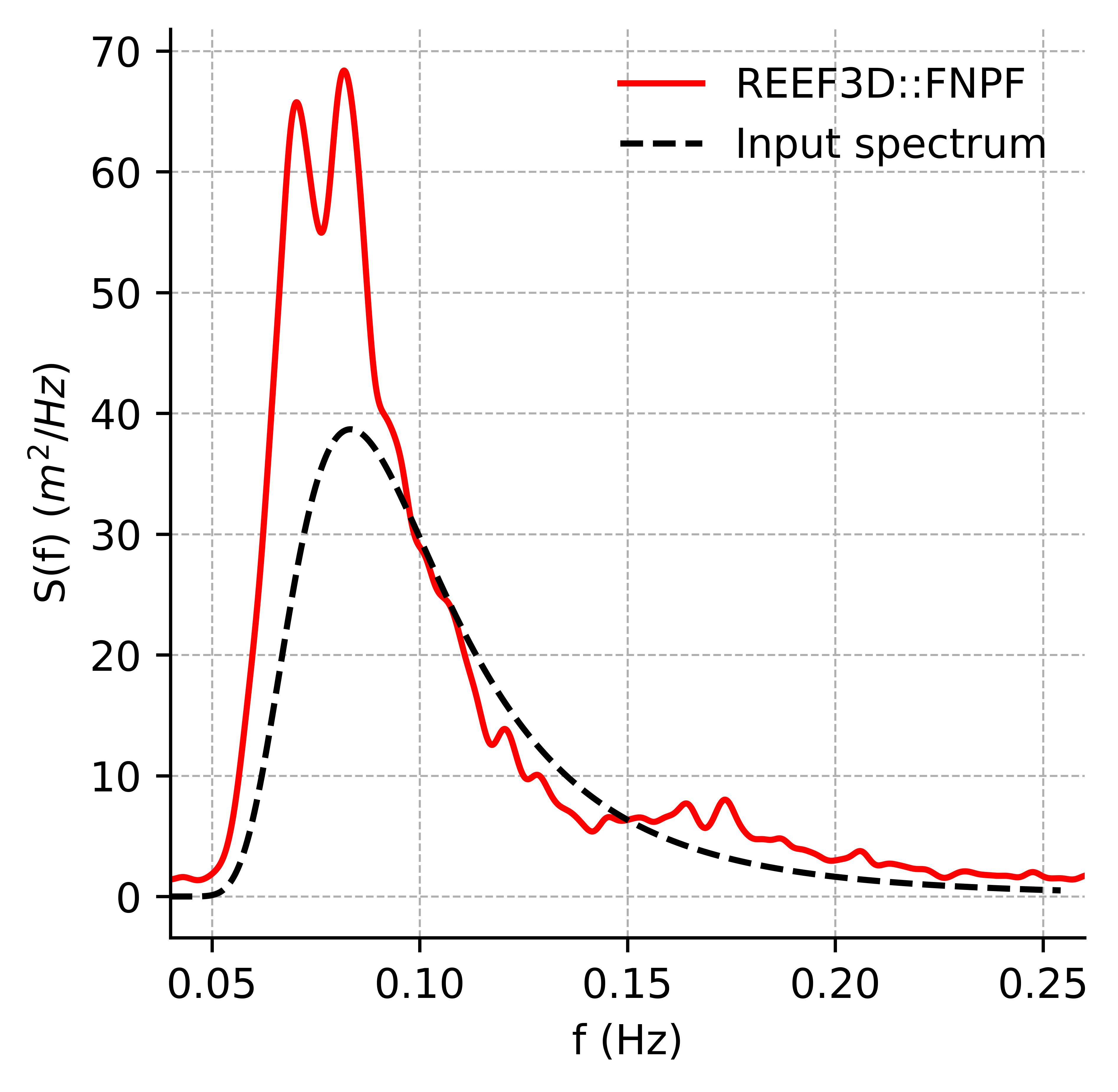}
                \caption{Navigation tower wave gauge $\triangle$}
                \label{fig:gauge_2}
\end{subfigure}
\begin{subfigure}[b]{0.3\textwidth}
                \centering
                \includegraphics[width=\textwidth]{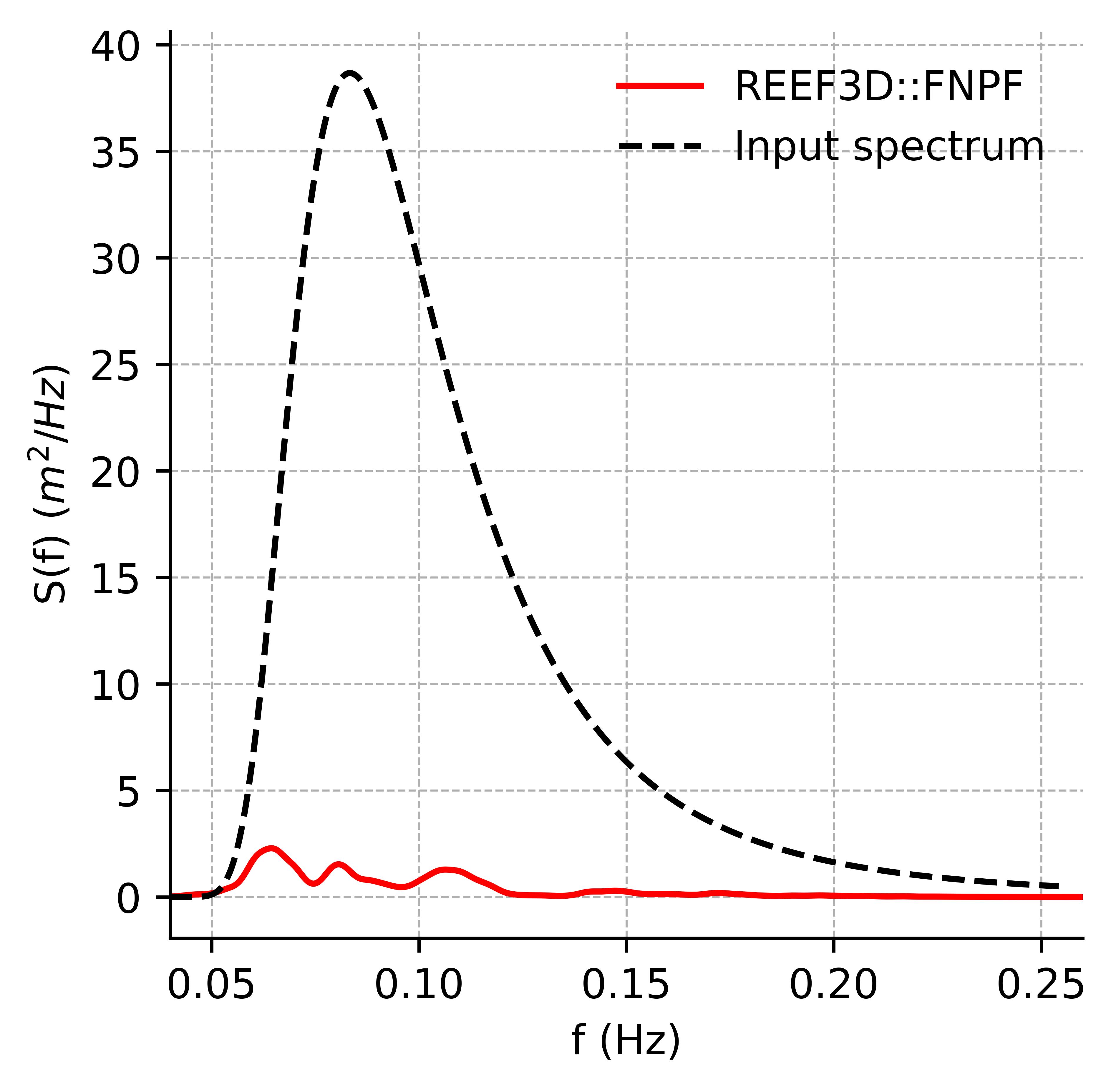}
                \caption{Anchorage wave gauge $\bigcirc$}
                \label{fig:gauge_3}
\end{subfigure}
\caption{The wave spectra obtained from the FFT analysis for the free surface elevation time record in the phase-resolving simulation with REEF3D::FNPF.}
\label{fig:spec_fnpf}
\end{figure}

The wave spectra recorded at the three wave gauges in REEF3D::FNPF are compared to the input wave spectrum in Fig.~\ref{fig:spec_fnpf}. The first wave gauge closely matches the input wave spectrum across all frequencies, with fluctuations reflecting the influence of varying bathymetry within the wave generation zone. At the navigation tower gauge, a pronounced energy increase near the peak frequency confirms shoaling effects. Conversely, the spectrum at the anchorage exhibits significantly reduced energy due to diffraction. However, discrepancies arise in the diffraction zone between phase-averaging and phase-resolving approaches.

\begin{figure}[!hb]	
\centering
\begin{subfigure}[b]{0.32\textwidth}
                \centering
                \includegraphics[width=\textwidth]{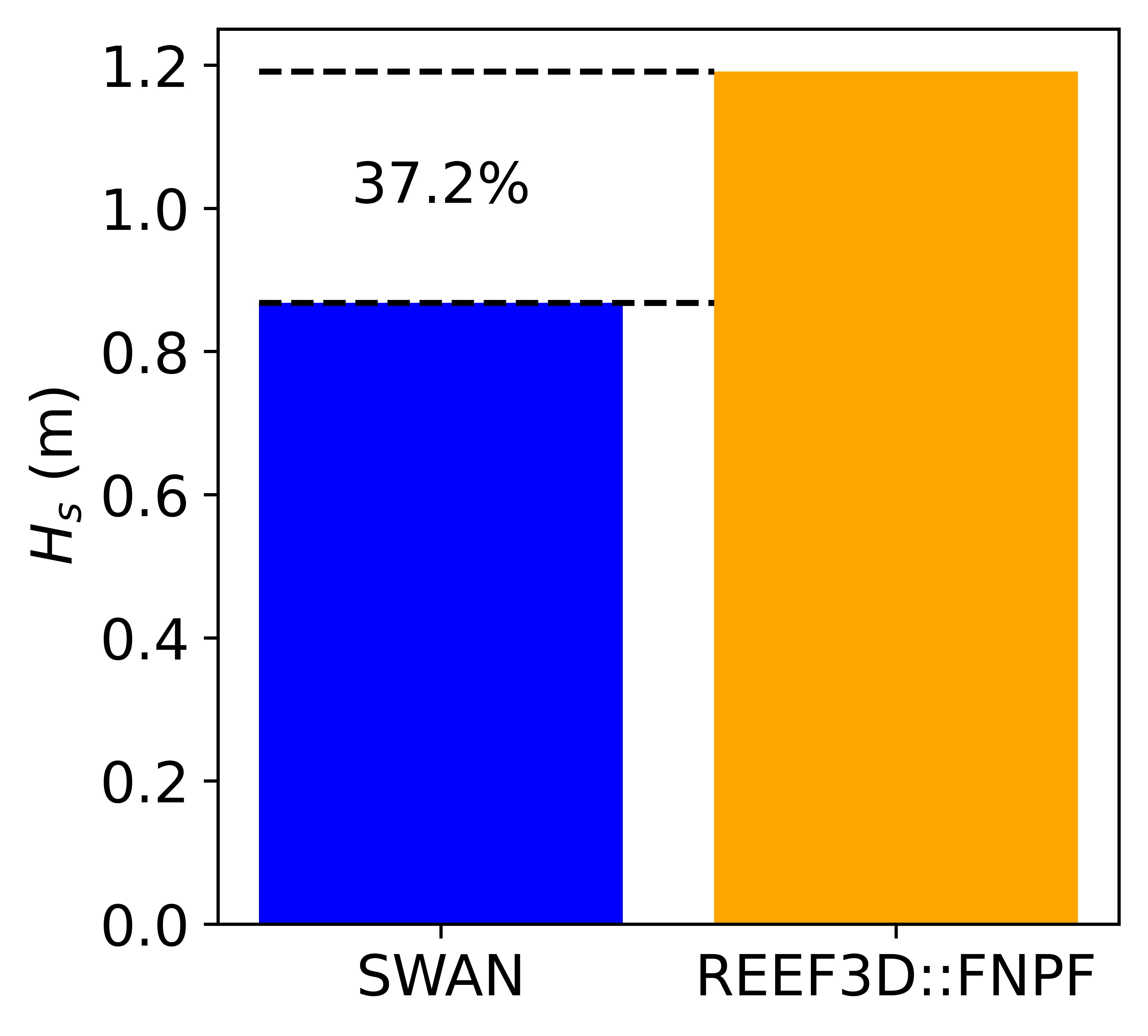}
                \caption{$H_s$}
                \label{fig:swan_comp}
\end{subfigure}
\begin{subfigure}[b]{0.53\textwidth}
                \centering
                \includegraphics[width=\textwidth]{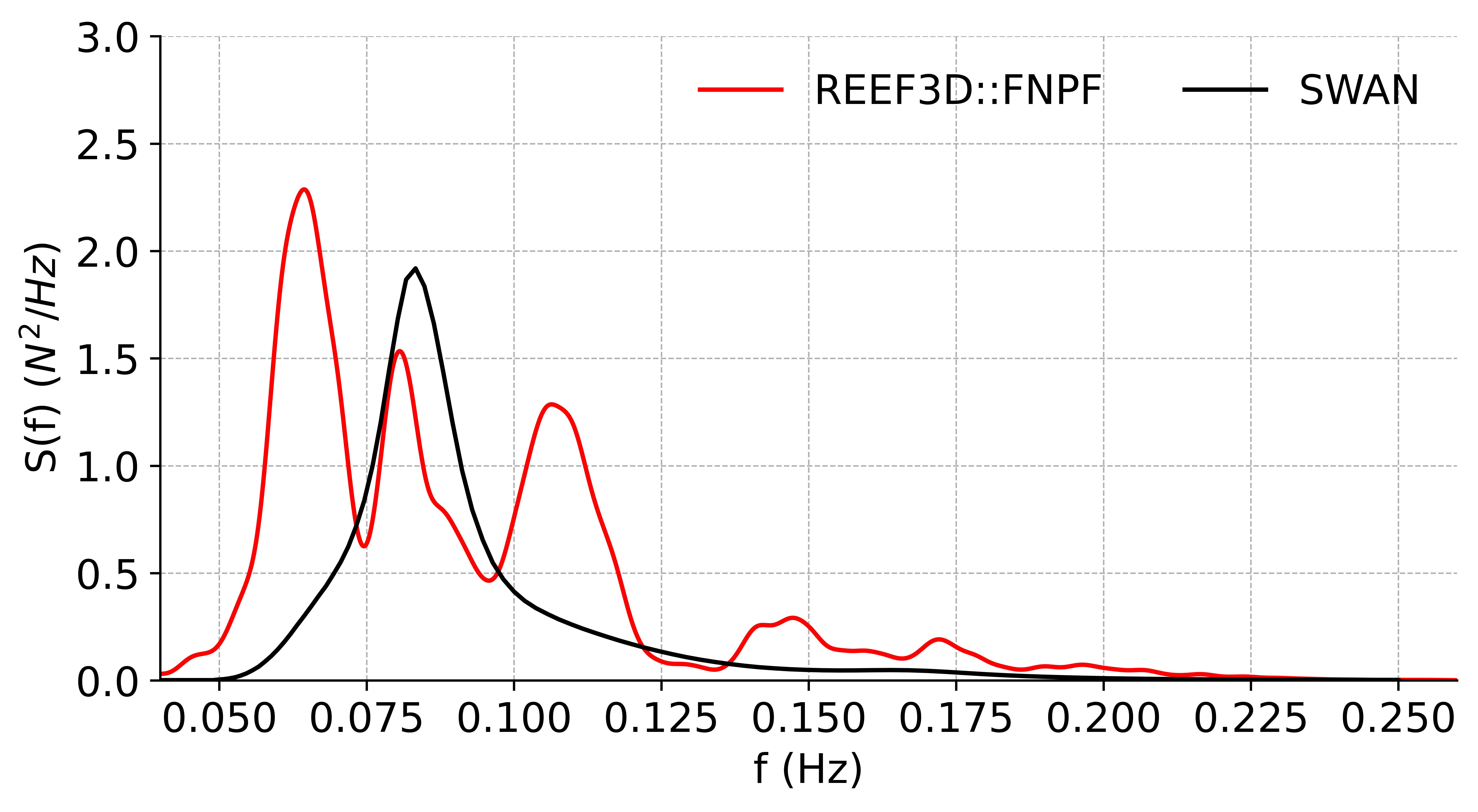}
                \caption{Wave spectra}
                \label{fig:fnpf_comp}
\end{subfigure}
\caption{The comparison of the $H_s$ and wave spectra at the anchorage from the phase-averaging and phase-resolving approach.}
\label{fig:compare}
\end{figure}

A comparison of $H_s$ and spectra at the anchorage between the two modeling approaches is presented in Fig.~\ref{fig:compare}. The phase-averaging approach underestimates $H_s$ by a notable percentage, whereas the phase-resolving method reveals a three-peak spectrum, indicating interactions between wave systems behind the island and nonlinear redistribution processes. This three-peak pattern is crucial for structural design and moored vessel safety, as it highlights potential resonance frequencies that could induce large motions. In contrast, the phase-averaging approach suggests only minor spectral modifications, resembling the input single-peak JONSWAP spectrum with reduced magnitude. For engineering applications, the phase-resolving model provides a more conservative wave condition estimate and identifies dominant wave frequencies, aiding in the design of structures to avoid resonance-related risks.

\begin{figure}[!hptb]
 	\centering
 	 	\includegraphics[width=0.75\textwidth]{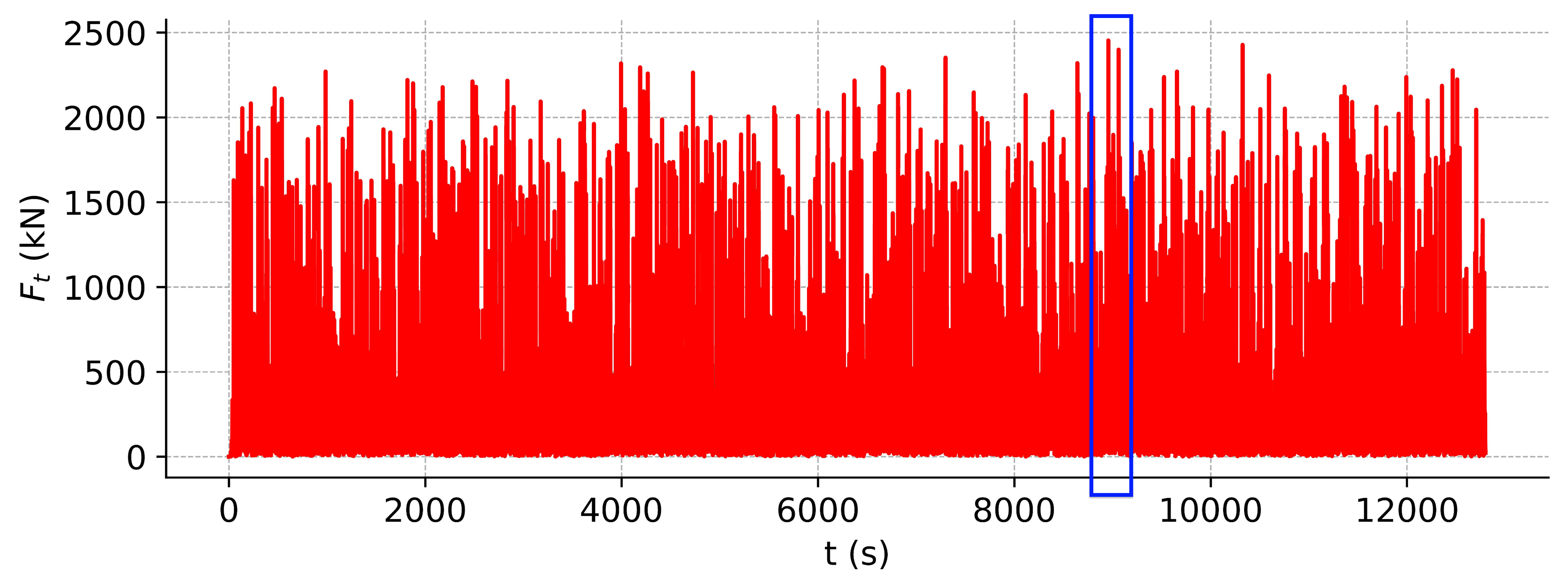}
 	\caption{Total forces on the navigation tower estimated with the ALE method in the REEF3D::FNPF NWT.} 
 	\label{fig::ale}
 \end{figure}

The conceptual navigation tower has a radius of 3 m, with an inertia coefficient of $C_M=1.5$ and a drag coefficient of $C_D=0.9$. The structure itself is not explicitly resolved within the FNPF NWT. Instead, the Arbitrary Lagrangian-Eulerian (ALE) force estimation method is used to compute wave loads in real time, utilizing undisturbed, fully nonlinear wave kinematics from the potential flow solver. The resulting total force time series is depicted in Fig.~\ref{fig::ale}. Unlike the free surface elevation time series, the total force time series directly reflects extreme load events by accounting for wave kinematics in addition to surface elevation effects.

\begin{figure}[!hptb]	
\centering
\begin{subfigure}[b]{0.3\textwidth}
                \centering
                \includegraphics[width=\textwidth]{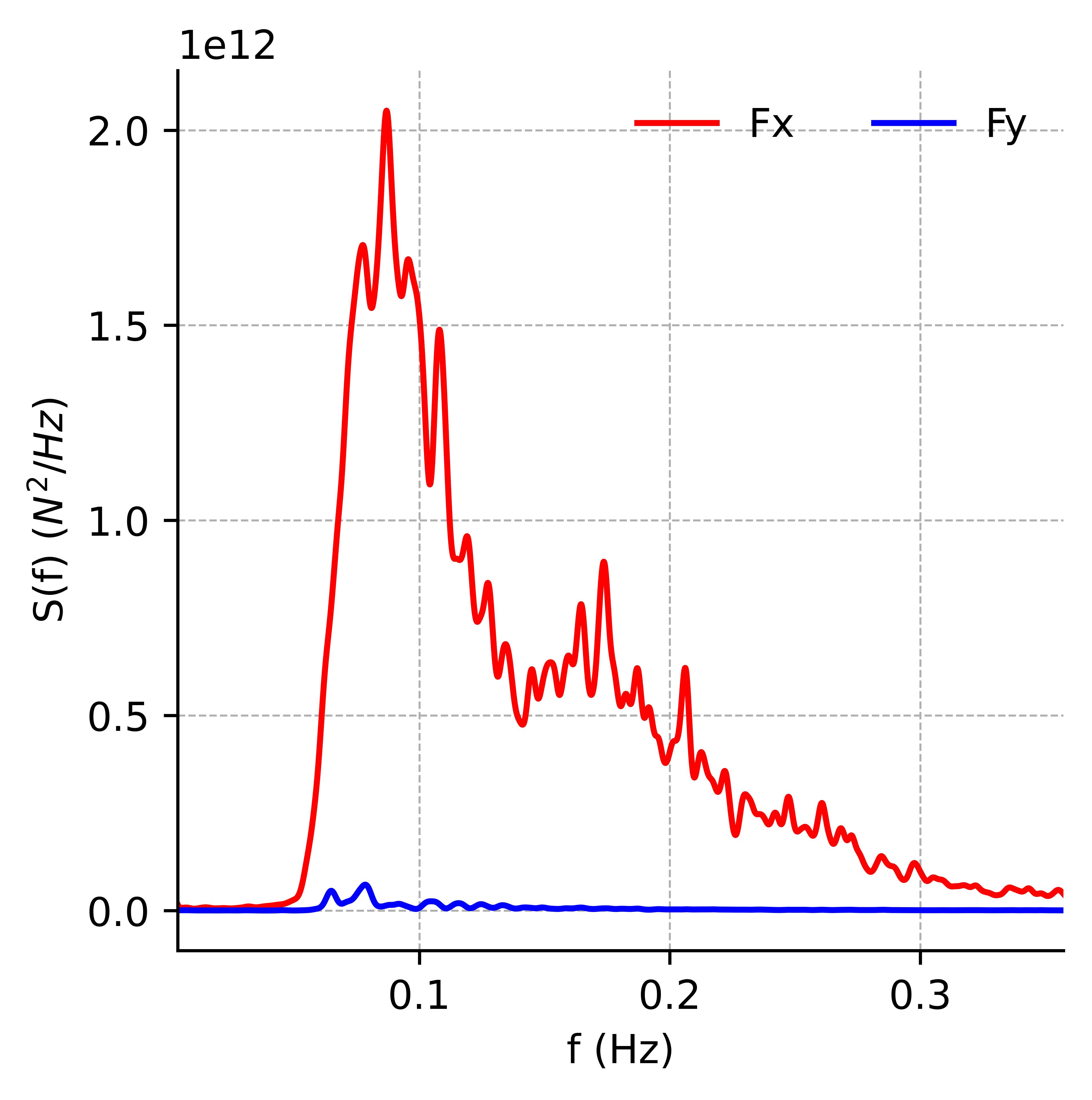}
                \caption{Force spectra in both directions}
                \label{fig:alex_xy}
\end{subfigure}
\begin{subfigure}[b]{0.3\textwidth}
                \centering
                \includegraphics[width=\textwidth]{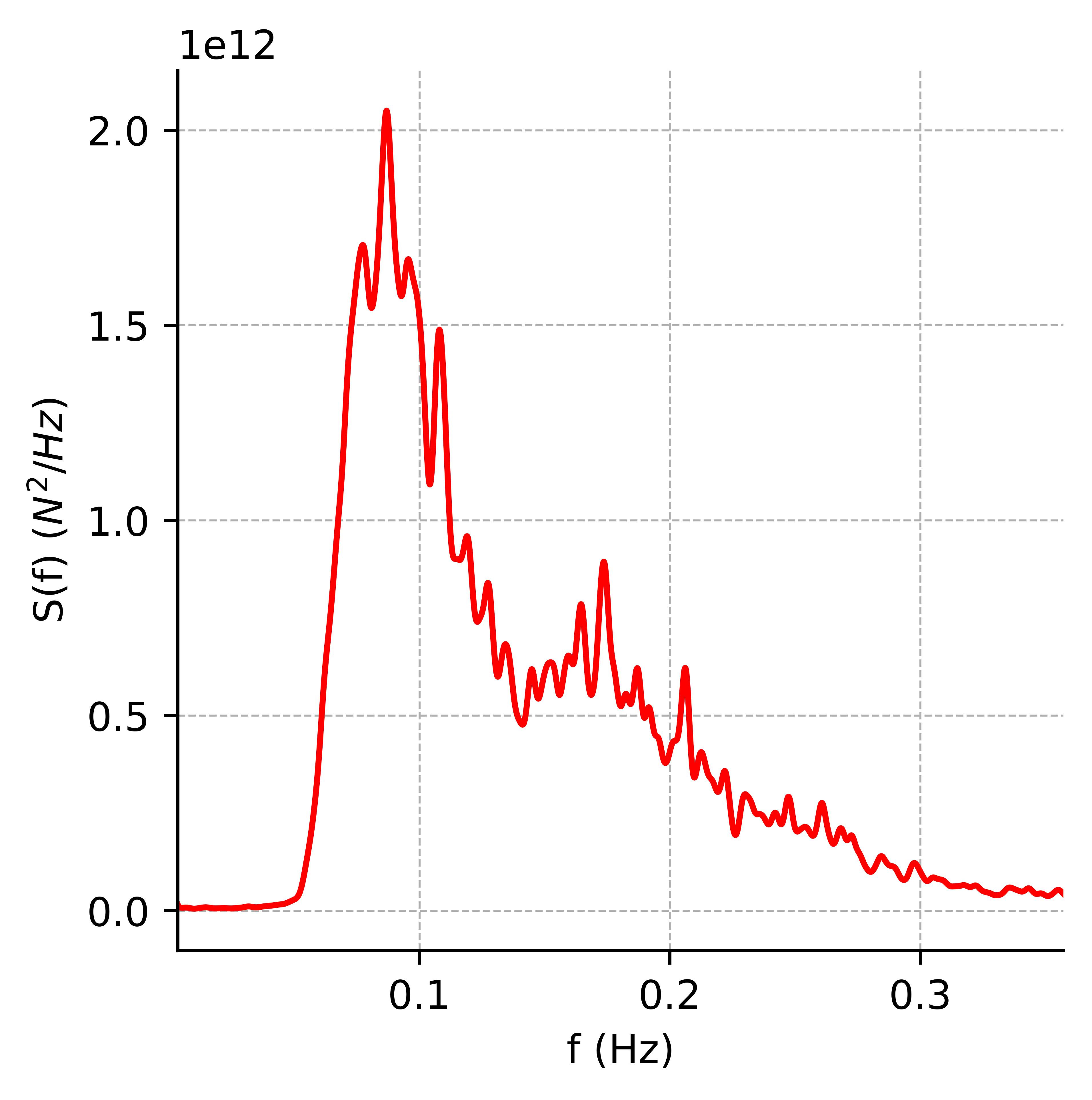}
                \caption{Force spectra in the x-direction}
                \label{fig:ale_x}
\end{subfigure}
\begin{subfigure}[b]{0.3\textwidth}
                \centering
                \includegraphics[width=\textwidth]{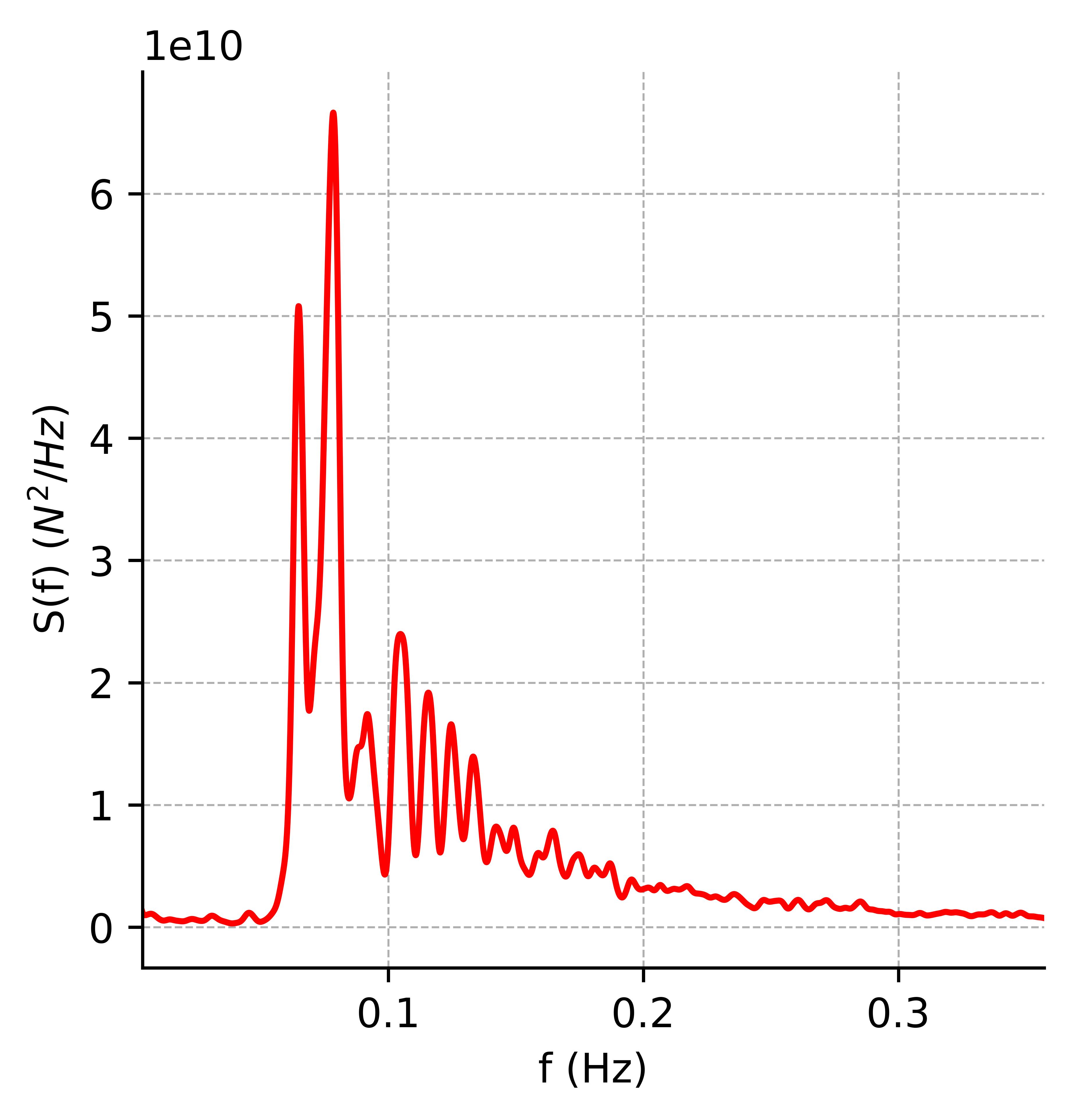}
                \caption{Force spectra in the y-direction}
                \label{fig:ale_y}
\end{subfigure}
\caption{Wave force spectra in the x and y directions at Lyngholmsboen. }
\label{fig:ale_spec}
\end{figure}

The spectra of the x- and y-components of the total force are shown in Fig.\ref{fig:ale_spec}. The force along the x-axis is significantly larger than that in the y-direction. Both spectra reveal a secondary peak at a higher frequency than the peak wave frequency, indicating the potential presence of high-frequency ringing. A time window from 8900 s to 9000 s, highlighted in Fig.\ref{fig::ale}, shows particularly large wave forces, likely due to wave breaking and slamming on the structure. To investigate this phenomenon in greater detail, this period is analyzed using the fully viscous CFD solver REEF3D::CFD.

To optimize computational efficiency, a near-field domain is selected for the CFD simulation, represented by the red box in the right panel of Fig.~\ref{fig:down-scale}. This localized approach significantly reduces the coupling data output from REEF3D::FNPF and improves the efficiency of grid interpolation between the FNPF and CFD domains. A Dirichlet boundary condition is applied to transfer wave kinematics and surface elevation data between the solvers in a one-way coupling approach.

The CFD simulation with REEF3D::CFD is conducted within a 150 × 30 m domain, discretized using a uniform cell size of 0.5 m, resulting in approximately 20.52 million computational cells. In this simulation, the navigation tower is fully resolved. A wave gauge is placed 6 m upstream from the tower center to capture the wave conditions. To estimate wave loads, a 6 × 6 m force box is centered around the tower, integrating pressure fields over the wetted surface.

The CFD simulation initializes flow conditions at 8900 s based on interpolated data from the FNPF domain and runs for 100 s. Due to the complexity of breaking waves, a strict CFL criterion of 0.1 is applied. The simulation is executed on the Betzy supercomputer, using 512 AMD EPYC 7742 cores (2.25 GHz) running on a Red Hat system, with a total computation time of 7.8 hours.

\begin{figure}[!hptb]	
\centering
\begin{subfigure}[b]{0.445\textwidth}
                \centering
                \includegraphics[width=\textwidth]{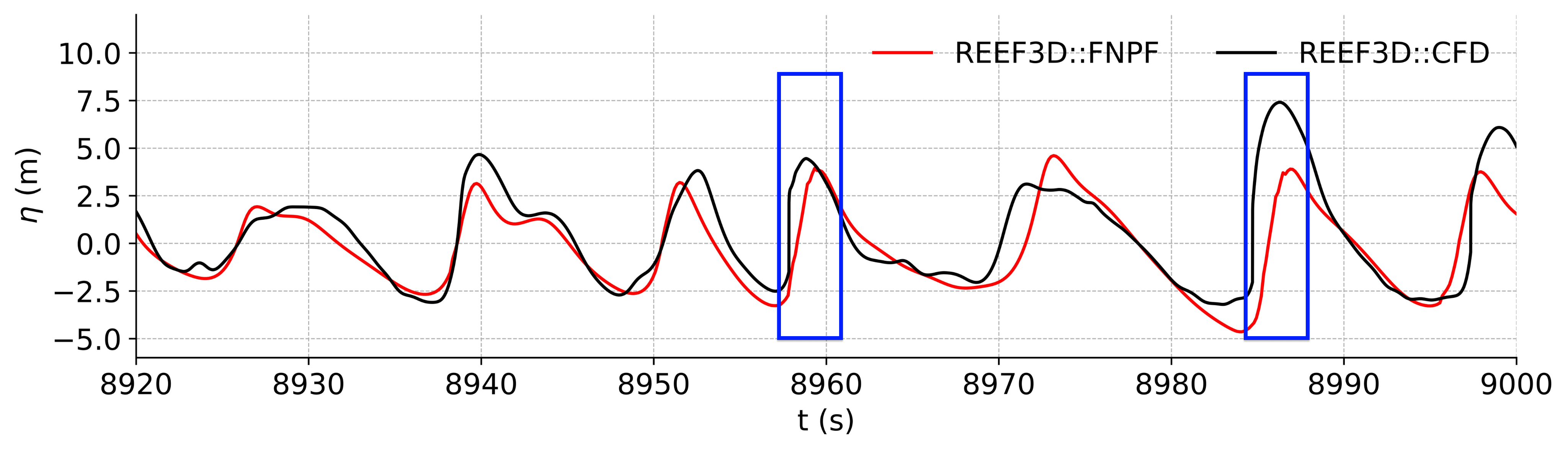}
                \caption{free surface elevation comparison }
                \label{fig:fsf_fnpf_cfd}
\end{subfigure}
\begin{subfigure}[b]{0.465\textwidth}
                \centering
                \includegraphics[width=\textwidth]{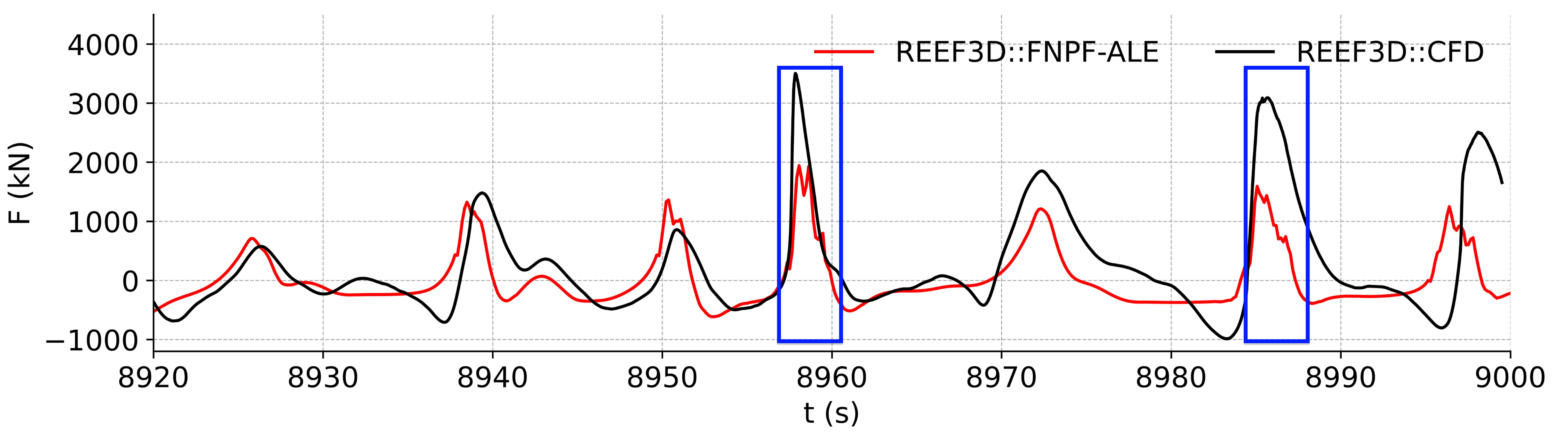}
                \caption{Wave force comparison }
                \label{fig:force_fnpf_cfd}
\end{subfigure}
\caption{Comparison of the free surface elevation and total wave force at the navigation tower between FNPF and HDC\&CFD simulations.}
\label{fig:hdc}
\end{figure}

\begin{figure}[!hptb]	
\centering
\begin{subfigure}[b]{0.465\textwidth}
                \centering
                \includegraphics[width=\textwidth]{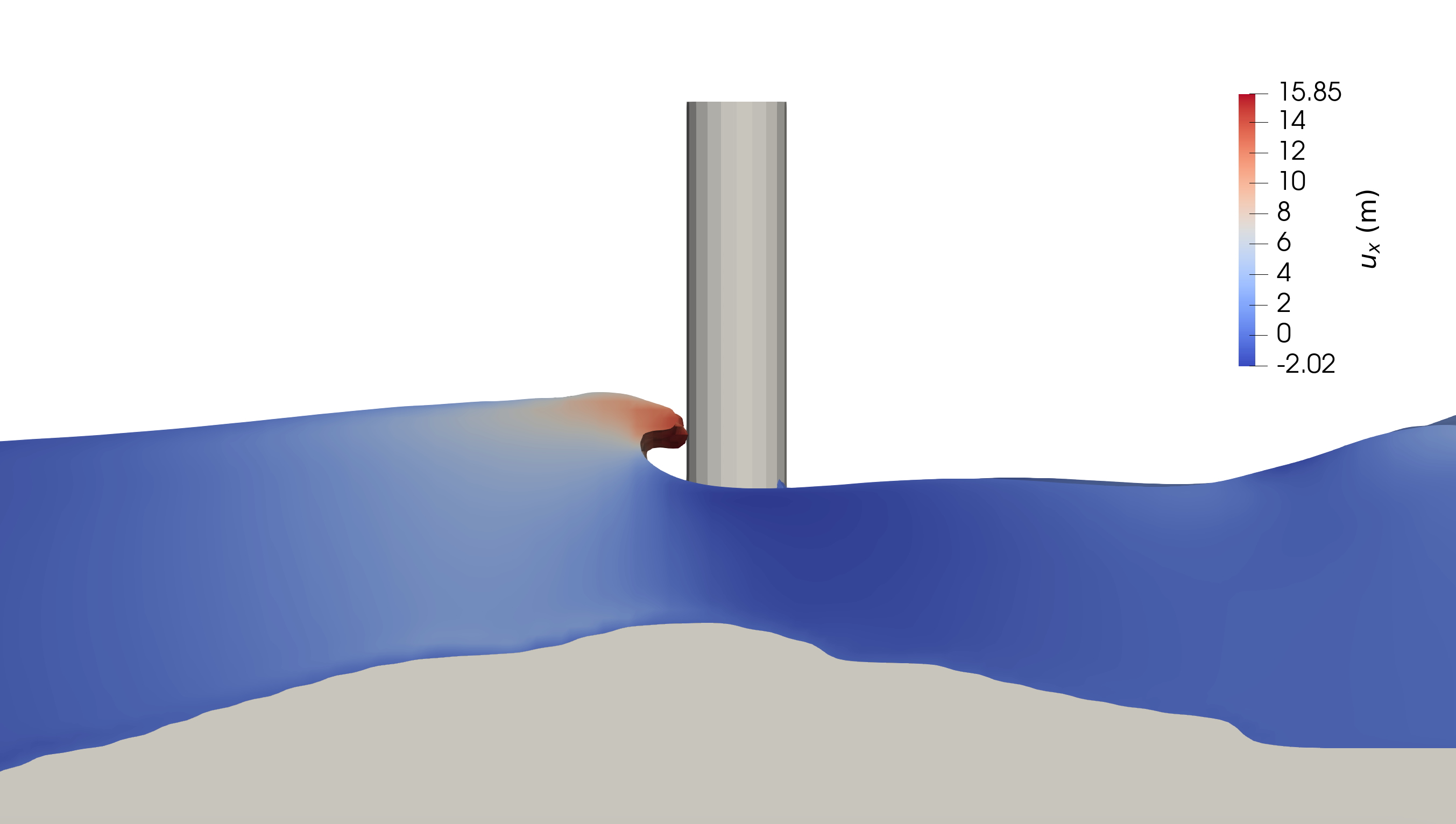}
                \caption{$t=8958.1$ s, side view}
                \label{1st_side}
\end{subfigure}
\begin{subfigure}[b]{0.465\textwidth}
                \centering
                \includegraphics[width=\textwidth]{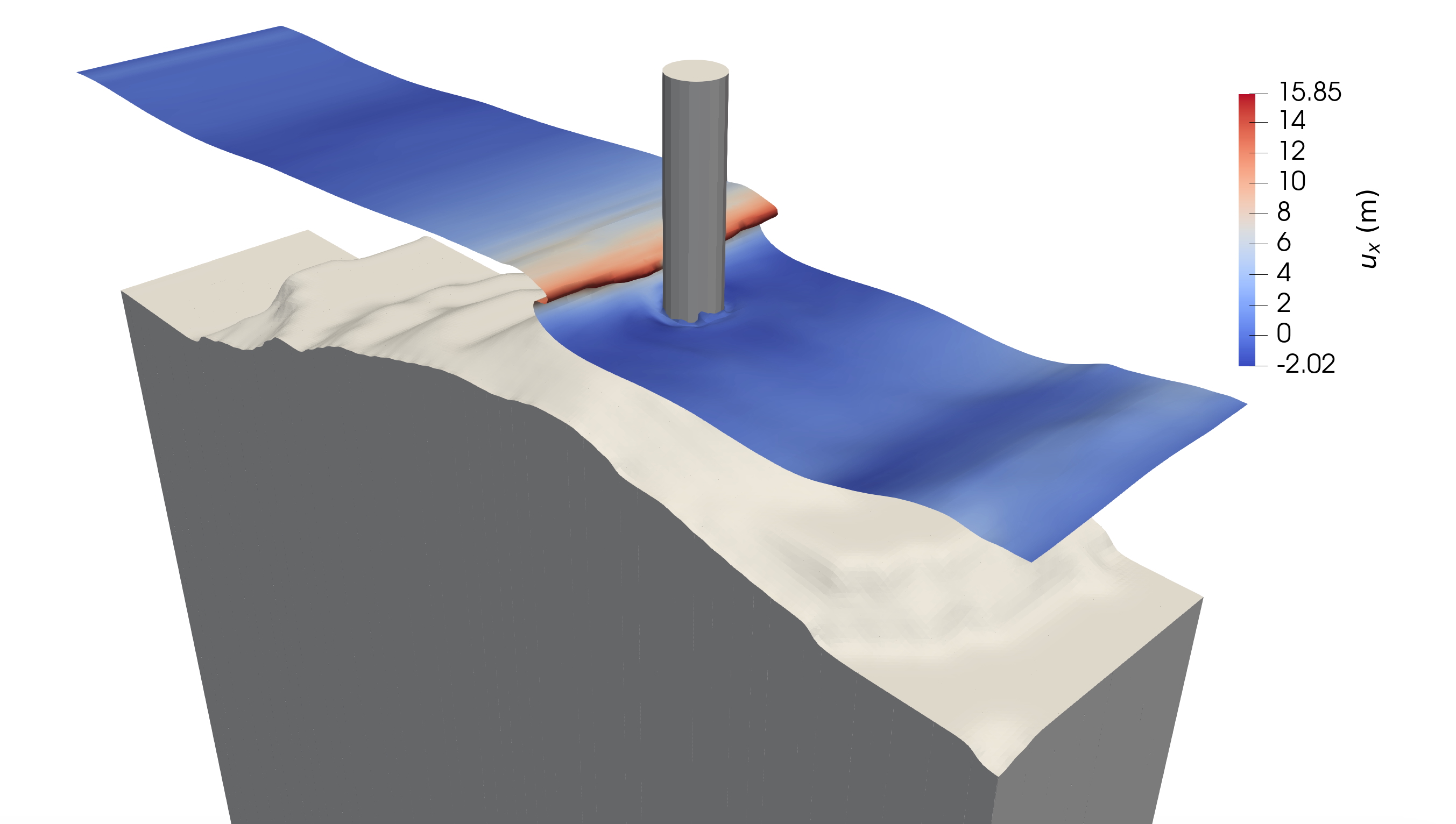}
                \caption{$t=8958.1$ s, top view}
                \label{1st_top}
\end{subfigure}
\begin{subfigure}[b]{0.465\textwidth}
                \centering
                \includegraphics[width=\textwidth]{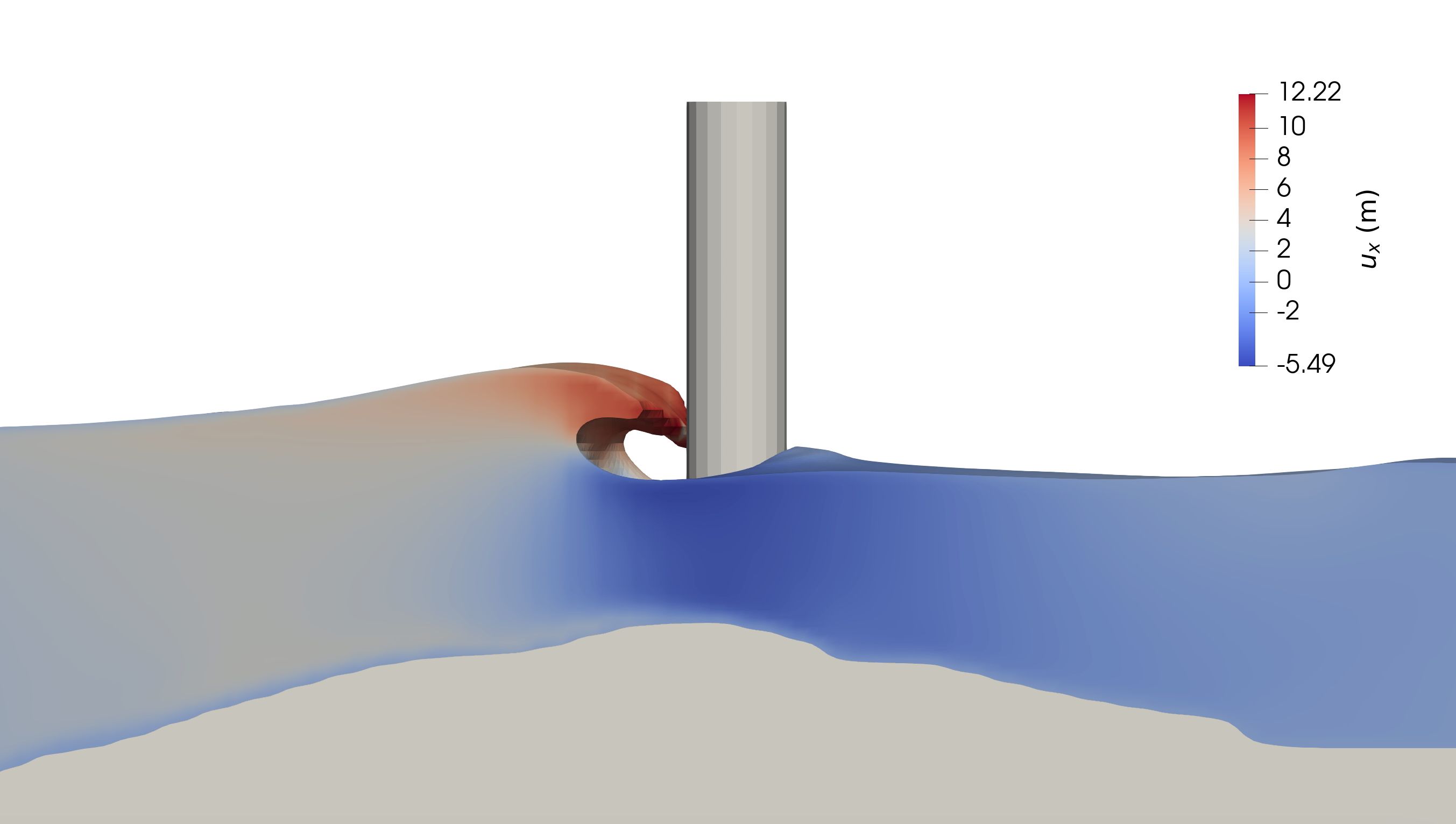}
                \caption{$t=8985.0$ s, side view}
                \label{2nd_side}
\end{subfigure}
\begin{subfigure}[b]{0.465\textwidth}
                \centering
                \includegraphics[width=\textwidth]{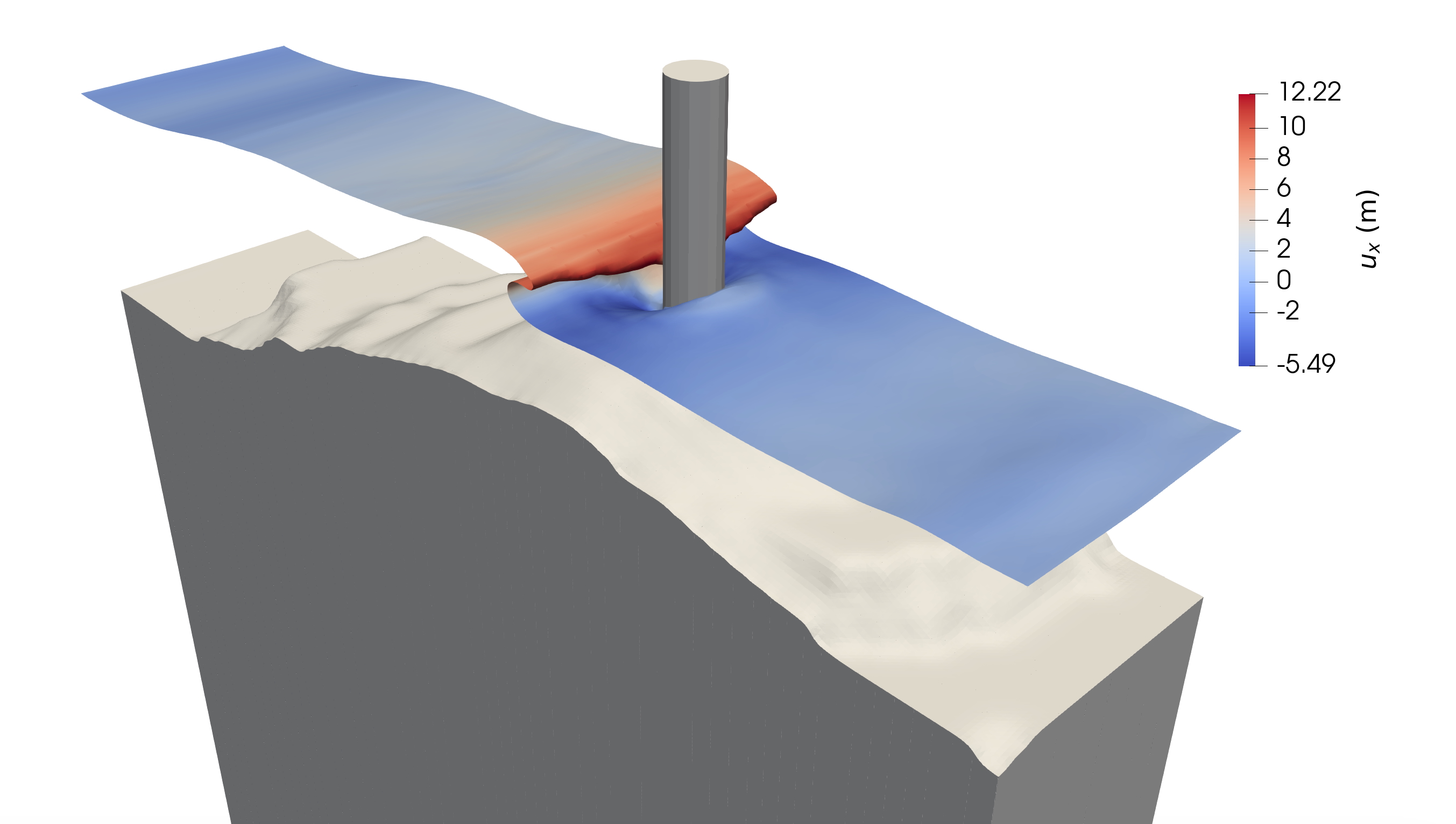}
                \caption{$t=8985.0$ s, top view}
                \label{2nd_top}
\end{subfigure}
\caption{The surface elevation rendered with the particle velocity magnitude at the two extreme wave load events at $t=8958.1$ s and $t=8985.0$ s in the CFD simulation through the HDC protocol. }
\label{fig:BR_CFD}
\end{figure}

The free surface elevation and wave forces computed through the HDC protocol in the CFD simulation are compared with the FNPF results in Fig.\ref{fig:hdc}. Two significant force events are identified in Fig.\ref{fig:force_fnpf_cfd}. The latter corresponds to a large wave crest, while the first extreme event occurs despite moderate wave crest heights in both the FNPF and CFD simulations. To investigate further, these extreme load events are visualized using ParaView at $t=8958.1$ s and $t=8985.0$ s, as shown in Fig.~\ref{fig:BR_CFD}. In both cases, plunging breaking waves impact the structure with high velocities, confirming slamming wave loads.

\begin{figure}[!hptb]	
\centering
\begin{subfigure}[b]{0.465\textwidth}
                \centering
                \includegraphics[width=\textwidth]{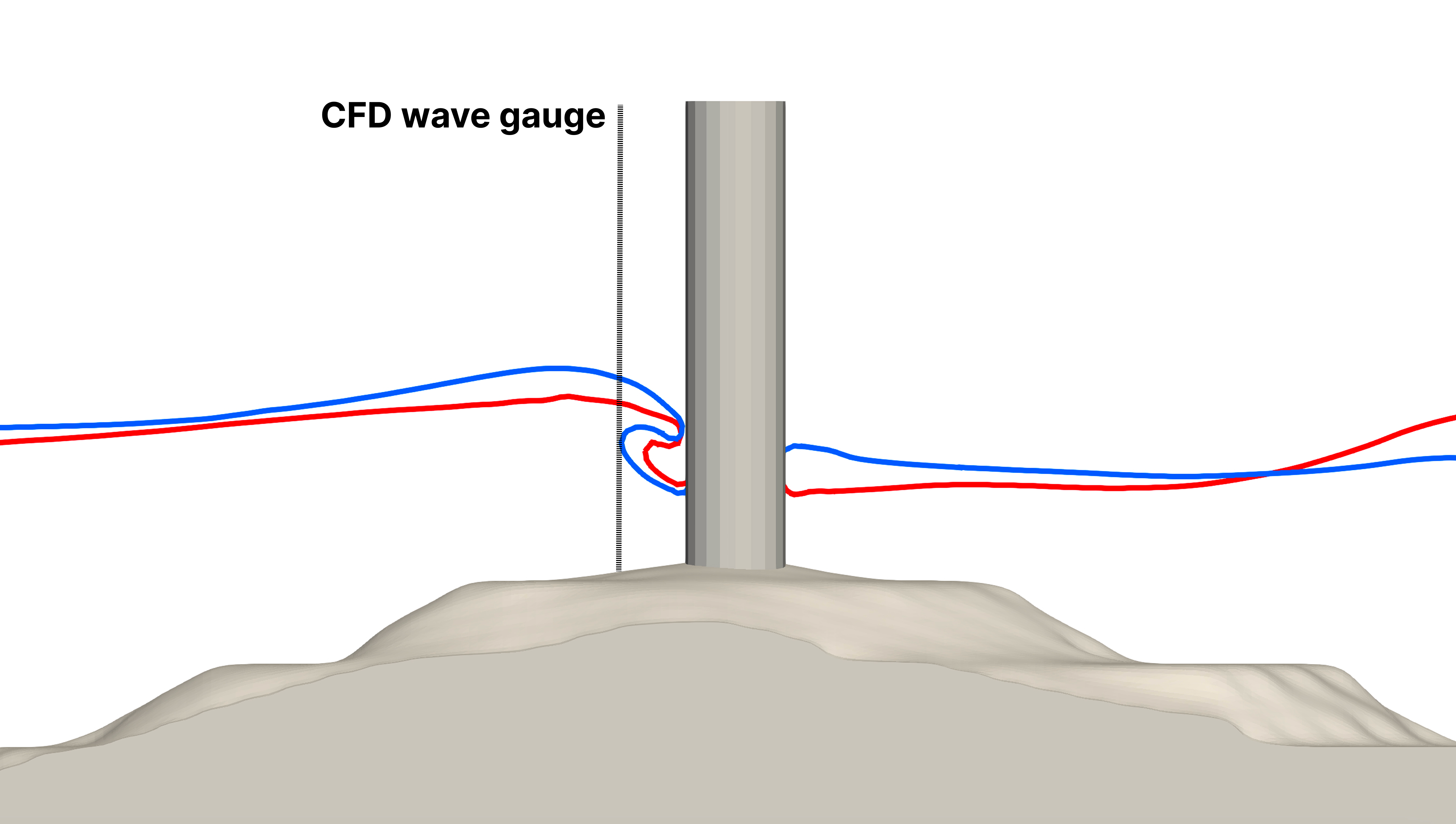}
                \caption{moment of breaking wave impact}
                \label{fig:fsf_br1}
\end{subfigure}
\begin{subfigure}[b]{0.465\textwidth}
                \centering
                \includegraphics[width=\textwidth]{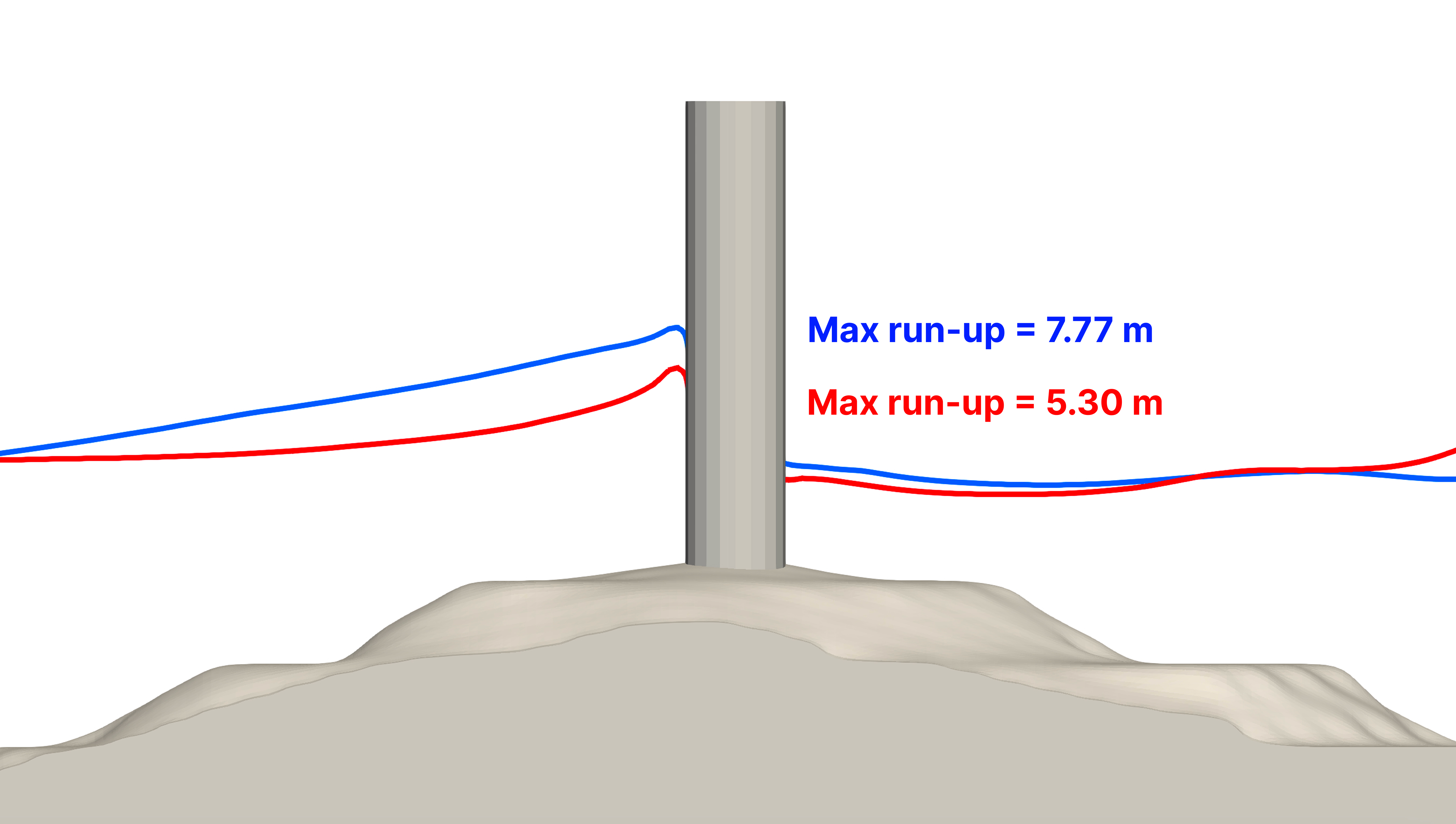}
                \caption{moment of maximum runup}
                \label{fig:fsf_br2}
\end{subfigure}
\caption{Comparison for the free surface elevation profiles simulated with REEF3D::CFD for the moment of breaking wave impacts and the moment of maximum runup at the two extreme events. The red lines represent the first extreme event at $t=8958.1$ s and the blue lines represent the second extreme event at $t=8985.0$ s. }
\label{fig:br_compare}
\end{figure}

A comparison of surface elevation profiles at these time points is presented in Fig.~\ref{fig:br_compare}. The first event ($t=8958.1$ s) exhibits a relatively lower wave crest height, consistent with free surface elevation observations. Similarly, maximum run-up levels are compared, showing that the second event at $t=8985.0$ s results in greater wetted surface area and larger run-up. However, the velocity magnitude during the first event reaches 15.85 m/s—approximately 30\% higher than the second event, where the maximum velocity is 12.22 m/s. This higher velocity magnitude is likely responsible for the increased wave force observed in the first event.

This analysis highlights that extreme wave forces, particularly slamming loads, are not solely correlated with wave crest height or surface elevation. Instead, wave breaking dynamics, shape, and velocity magnitude play crucial roles. The HDC protocol and CFD simulation provide the necessary flow details to accurately capture and explain these hydrodynamic phenomena, offering valuable insights for structural design and wave impact assessments.

The time efficiency of the entire NORA-SARAH down-scale wave modelling approach is summarized in Table~\ref{tab:time}. The approach combines the strengths of different model types, switching scales with the consideration of the level of resolved physical details and computational efficiency. As a result, the entire process can be performed within 11 hours. 

\begin{table}[!hptb]
\begin{tabular}{l|lll}
\hline
 & SWAN & REEF3D::FNPF & REEF3D::CFD \\ \hline
Simulation time & stationary & 3.6 hours & 100 seconds \\
Elapsed time & 0.7 hours & 4.5 hours & 7.8 hours  \\
Computational resource  & 4 $\times$ M4 Max & 128 $\times$ AMD & 512 $\times$ AMD \\ \hline
\end{tabular}
\caption{Simulation time and elapsed time during the NORA-SARAH down-scale wave modeling procedure.}
\label{tab:time}
\end{table}


\section{Conclusion}
The NORA-SARAH approach (\textbf{NORA}-\textbf{S}W\textbf{A}N-\textbf{R}EEF3D-\textbf{A}LE-\textbf{H}DC) introduces an advanced down-scaling procedure for hydrodynamic analysis, effectively bridging the gap between large-scale offshore metocean data and site-specific coastal engineering applications over varying bathymetric conditions. By integrating multiple numerical wave models at different scales, each with varying levels of physical complexity, the approach ensures minimal information loss while delivering detailed insights into wave transformation and wave loads under relevant hydrodynamic events. This multi-scale modeling strategy strikes a balance between computational efficiency and accuracy, making it a practical and robust tool for coastal engineering applications.

The framework leverages the open-access hindcast database NORA3, the open-source phase-averaging model SWAN, and the open-source hydrodynamic model REEF3D. Within this structure, REEF3D's comprehensive coastal wave suite enables high-resolution wave transformations through a fully nonlinear potential flow model, facilitates extreme event screening via ALE force estimation, and provides high-fidelity CFD simulations to resolve complex wave-breaking and flow interactions. The seamless integration of non-viscous and viscous solvers over irregular bathymetry showcases a strong hydrodynamic coupling (HDC) capability, further enhancing the model’s applicability in real-world scenarios.

A case study in southern Norway demonstrates the effectiveness of the NORA-SARAH approach, analyzing slamming wave loads on a navigation tower and wave diffraction at an anchorage. The findings highlight key aspects of wave modeling, including the tendency of phase-averaged methods to underpredict wave diffraction compared to phase-resolving approaches. Moreover, the study reinforces the notion that wave height alone is not a sufficient predictor of wave loads—wave kinematics play a crucial role, particularly in breaking wave impacts. The integration of ALE extreme event identification (EEI) with HDC analysis emphasizes the necessity of the complete down-scaling process, while also showcasing significant time efficiency gains, as only a limited domain and time window require CFD-level analysis. Furthermore, the framework’s open accessibility provides a valuable blueprint for adaptation, fostering future developments toward automation. When fully automated, the NORA-SARAH framework holds the potential to become an integral component of data-driven design optimization and serve as a foundation for digital twin applications.

Compared to traditional $H_s$ and spectrum-based design methodologies, the NORA-SARAH approach offers several distinct advantages: 
\begin{itemize}
\item Better-resolved diffraction; 
\item Fast extreme wave load identification; 
\item Detailed breaking wave structures and near-field hydrodynamics
\item Balance between computational efficiency and the level of details and accuracy 
\item An open and streamlined framework for easy adaptation and automation
\end{itemize}

In the present study, the input waves for both phase-averaging and phase-resolving models are parametrized using $H_s$ and $T_p$ from the upper-level scale, besides the HDC within REEF3D. However, ongoing developments aim to enhance the framework by directly transferring both frequency and directional spectra across the down-scaling process and among numerical models. Additionally, future enhancements include modular batch boundary conditions for multi-stressor, multi-system simulations. These improvements will enable the analysis of interacting wave systems from different principal directions and incorporate the effects of currents and wind on wave fields, further strengthening the applicability and accuracy of the proposed methodology.

The NORA-SARAH approach represents a significant step forward in high-fidelity coastal engineering modeling. Its ability to integrate multi-scale hydrodynamic processes while maintaining computational efficiency makes it a valuable tool for engineers and researchers alike. With continued development, this open framework has the potential to revolutionize the way wave-induced hydrodynamic forces are analyzed and optimized in coastal and offshore environments.


\section*{Acknowledgments}
The simulations were performed on the supercomputer Betzy provided by UNINETT Sigma2 - the National Infrastructure for High-Performance Computing and Data Storage in Norway. The authors also thank the funding by the European Union (ERC, PARTRES, 101045646). Views and opinions expressed are however those of the authors only and do not necessarily reflect those of the European Union or the European Research Council Executive Agency. Neither the European Union nor the granting authority can be held responsible for them.

\bibliographystyle{elsarticle-num-names}
\bibliography{All_in_One_Tidy}

\end{document}